\documentclass{article}

\usepackage[preprint]{neurips_2026}

\usepackage[utf8]{inputenc} 
\usepackage[T1]{fontenc}    
\usepackage{hyperref}       
\usepackage{url}            
\usepackage{booktabs}       
\usepackage{amsfonts}       
\usepackage{nicefrac}       
\usepackage{microtype}      
\usepackage{xcolor}         

\usepackage[dvipsnames]{xcolor}
\usepackage{multirow}
\usepackage{amsmath}
\usepackage{mathtools}
\usepackage{braket}
\usepackage{changes}
\definechangesauthor[name={kien}, color=blue]{kn}
\definechangesauthor[name={Ankit}, color=purple]{ak}

\title{QAP-Router: Tackling Qubit Routing as Dynamic Quadratic Assignment with Reinforcement Learning}

\author{%
  Kien X. Nguyen$^{1,2}$, Ankit Kulshrestha$^2$, Ilya Safro$^1$, Xiaoyuan Liu$^2$ \\
  $^1$Department of Computer and Information Sciences, University of Delaware, USA\\
  $^2$Quantum Lab, Fujitsu Research of America, USA\\
  \texttt{\{kxnguyen,isafro\}@udel.edu}, \texttt{\{akulshrestha,xliu\}@fujitsu.com} \\
}

\begin{document}

\maketitle
\begin{abstract}
  Qubit routing is a fundamental problem in quantum compilation, known to be NP-hard.
  Its dynamic nature makes local routing decisions propagate and compound over time, making global efficient solutions challenging.
  Existing heuristic methods rely on local rules with limited lookahead, while recent learning-based approaches often treat routing as a generic sequential decision problem without fully exploiting its underlying structure.
  In this paper, we introduce \texttt{QAP-Router}, framing qubit routing based on a dynamic Quadratic Assignment Problem (QAP) formulation.
  By modeling logical interactions, or quantum gates, as flow matrices and hardware topology as a distance matrix, our approach captures the interaction-distance coupling in a unified objective, which defines the reward in the reinforcement learning environment.
  To further exploit this structure, the policy network employs a solution-aware Transformer backbone that encodes the interaction between the flow matrix and the distance matrix into the attention mechanism.
  We also integrate a lookahead mechanism that blends naturally into the QAP framework, preventing myopic decisions.
  Extensive experiments on 1,831 real-world quantum circuits from the MQTBench, AgentQ and QUEKO datasets show that our method substantially reduces the CNOT gate count of routed circuits by \textbf{15.7\%}, \textbf{30.4\%} and \textbf{12.1\%}, respectively, relative to existing industry compilers.

\end{abstract}

\section{Introduction}
Quantum computing has seen tremendous progress over the past decade due to its vast potential applications, such as security~\cite{portmann2022security,li2023post}, optimization~\cite{farhi2014quantum,abbas2024challenges}, network science \cite{shaydulin2019network}, machine learning~\cite{biamonte2017quantum,cerezo2022challenges,zeguendry2023quantum,zhao2026exponential}, and
finance \cite{herman2023quantum}. 
These advances have enabled the execution of small- to medium-scale quantum circuits on real quantum hardware, which is now referred to as noisy intermediate-scale quantum (NISQ) systems.
For instance, in 2023, IBM demonstrated that its 127-qubit Eagle processor can execute quantum circuits beyond the reach of classical exact simulation, marking a milestone towards quantum utility~\cite{kim2023evidence,kremer2024practical}.
In 2025, Fujitsu and RIKEN developed a 256-qubit superconducting quantum computer, marking a significant step towards large-scale quantum systems~\cite{fujitsu256qubit}.

\begin{figure}
    \centering
    \includegraphics[width=0.85\linewidth]{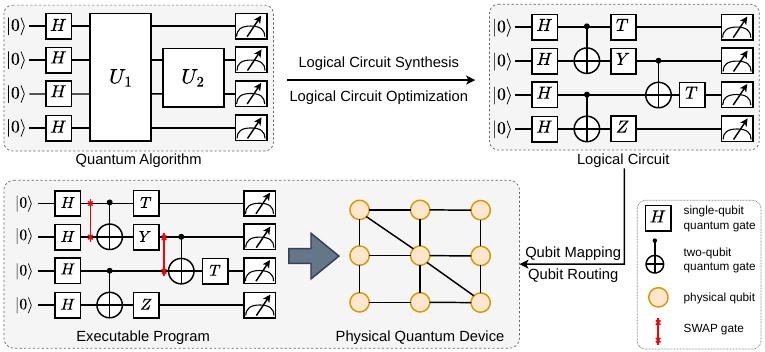}
    \caption{Overview of the quantum compilation pipeline. A quantum algorithm, expressed as a sequence of unitary transformations, is first translated into a logical circuit via circuit synthesis and optimization. Then, to execute the logical circuit on a target quantum device, qubit mapping determines an initial assignment of logical to physical qubits, followed by qubit routing that inserts SWAP operators to enable the execution of all two-qubit gates under the connectivity constraints.}
    \label{fig:pipeline}
\end{figure}

Despite this progress, the current NISQ hardware, containing tens to a few hundred qubits, remains highly constrained.
They often support only a restricted set of native gates and have limited qubit connectivity, so arbitrary qubit pairs cannot directly interact.
Therefore, quantum algorithms must be transformed into equivalent circuits that comply with device constraints through a process known as \textit{quantum compilation}~\cite{maronese2022quantum,yan2024quantum}.
This process contains two phases: (i) circuit synthesis and optimization, and (ii) qubit mapping and routing (Figure~\ref{fig:pipeline}).
In the first phase, a high-level quantum algorithm is translated into a gate-level circuit over the native instruction set of the target hardware, while compiler optimizations reduce depth and gate count~\cite{zulehner2018efficient,li2019tackling,sivarajah2021t,karuppasamy2025comprehensive}.
In the second phase, the logical circuit is compiled to the device constraints; qubit mapping creates an initial logical-to-physical assignment, and qubit routing inserts additional operations, such as SWAP gates, to make all two-qubit gates executable under the device connectivity~\cite{cowtan2019qubit,niu2020hardware,ushijima2023ising}.
Moreover, quantum operations are inherently noisy due to hardware imperfections and environmental disturbances that cause gate errors, decoherence, and readout noise. Therefore, it is crucial to keep circuits as shallow as possible by minimizing the number of inserted SWAP operations, which increase depth and accumulated errors.

In this paper, we focus on qubit routing on superconducting quantum devices, an NP-hard problem~\cite{ito2023algorithmic}, with the assumption that an initial mapping is given.
Recent work includes classical optimization methods and machine learning based approaches.
Classical methods include exact solutions~\cite{siraichi2018qubit}, heuristics~\cite{zulehner2018efficient,cowtan2019qubit,li2019tackling,zou2024lightsabre,bach2025efficient}, and reduction~\cite{murali2019noise,molavi2022qubit} to construct efficient mapping and routing strategies.
In contrast, machine learning approaches aim to learn a general routing policy that can be generalized across circuits and device topologies, typically by means of reinforcement learning and learned heuristics~\cite{herbert2018using,huang2022reinforcement,pozzi2022using,sinha2022qubit,tang2024alpharouter,kremer2024practical}.

However, existing approaches to qubit routing remain limited in their ability to produce globally efficient compilations.
In practice, routing decisions are often guided by heuristic objectives with limited lookahead, which can improve executability at the current stage of the circuit but may still introduce overhead.
As a result, qubit routing is rarely performed as an isolated step; instead, it is commonly interleaved with further compilation procedures that refine the circuit, adjust the mapping, or reduce the overhead introduced during routing~\cite{zulehner2018efficient,li2019tackling,kremer2024practical}.


To this end, we exploit the inherent structure of the qubit routing problem by framing it as \textit{a dynamic global assignment task}, where the logical circuit induces structured interactions, and the device topology defines communication costs.
Specifically, we propose a formulation of qubit routing based on the Quadratic Assignment Problem (QAP)~\cite{lawler1963quadratic,lee2004generalized} and use its objective to define the reward function in our reinforcement learning framework.
By representing logical qubit interactions as a flow matrix and hardware connectivity as a distance matrix, our approach directly optimizes the alignment between circuit structure and device topology, which we incorporate into the learning objective.
Furthermore, our policy network captures this interaction–distance coupling through a solution-aware Transformer encoder that jointly encodes flow-based interactions and topology-aware distances within the attention mechanism.~\cite{tan2024learning}.
This provides a principled foundation that strikes a good balance between optimization and learning, facilitating globally informed routing strategies.

Finally, we evaluate our approach on realistic quantum circuits against state-of-the-art qubit compilers and show that it achieves up to \textbf{15.7\%} reduction on the MQTBench dataset~\cite{quetschlich2023mqtbench}, \textbf{30.4\%} reduction on AgentQ dataset~\cite{jern2025agent}, and \textbf{12.1\%} reduction on QUEKO dataset~\cite{tan2020optimality} in CNOT gate count, in comparison to existing approaches.

In summary, our contributions are threefold:
\begin{enumerate}
    \item We introduce \texttt{QAP-Router}, framing qubit routing as a dynamic QAP and incorporating its objective into the reward function of our reinforcement learning framework.
    \item We design a solution-aware state encoder based on a Transformer backbone, which integrates the interaction between the flow matrix and the distance matrix into the attention mechanism.
    \item We train our method on randomly generated circuits and evaluate it on realistic benchmark circuits against existing commercial quantum compilers, showing that it achieves improved routing performance on three different datasets.
\end{enumerate}
\section{Background and Motivation}

\subsection{Quadratic Assignment Problem}
The Quadratic Assignment Problem (QAP) involves $N$ facilities and $M$ locations~\cite{lawler1963quadratic,lee2004generalized}.
A distance is specified for each pair of locations, and a weight (or flow) is specified for each pair of facilities.
The goal is to assign all facilities to locations so that the sum of distances multiplied by the corresponding flow is minimized.
Let $\boldsymbol{X}\in \mathbb{R}^{N\times M}$ be the permutation matrix, $\boldsymbol{F}\in\mathbb{R}^{N\times N}$ be the flow matrix, and $\boldsymbol{D}\in\mathbb{R}^{M\times M}$ be the distance matrix.
The objective function of QAP is
\begin{align}
    &\min_{\boldsymbol{X}} \mathrm{Tr}(\boldsymbol{F}\cdot \boldsymbol{X}\cdot \boldsymbol{D}\cdot \boldsymbol{X}^\top),\\
    \text{s.t. } &\sum_i \boldsymbol{X}_{ij} = 1, \quad\forall j,\quad\sum_j \boldsymbol{X}_{ij} = 1, \quad\forall i,\quad \boldsymbol{X}_{ij}\in\{0,1\},
\end{align}
where $\mathrm{Tr}(\cdot)$ denotes the trace of a matrix.
QAP has been studied extensively in the High-Performance Computing (HPC) field~\cite{hoefler2011generic,sudheer2012optimization,schulz2017better}.
In this work, we draw inspiration from the topology-aware process mapping in HPC, where the goal is to assign communicating processes (facilities) to physical processors (locations) so that frequently communicating processes are placed close to each other in the machine topology.
This is analogous to our qubit routing setting, where logical qubits play the role of communicating processes and physical qubits are the physical processors.
The main distinction is that HPC process mapping is often static, whereas qubit routing is dynamic, as the mapping changes over time through SWAP actions, and the interaction matrix evolves as gates are executed and cleared.
This motivates framing our problem to a dynamic QAP framework.

\begin{figure}[hb]
    \centering
    \includegraphics[width=\linewidth]{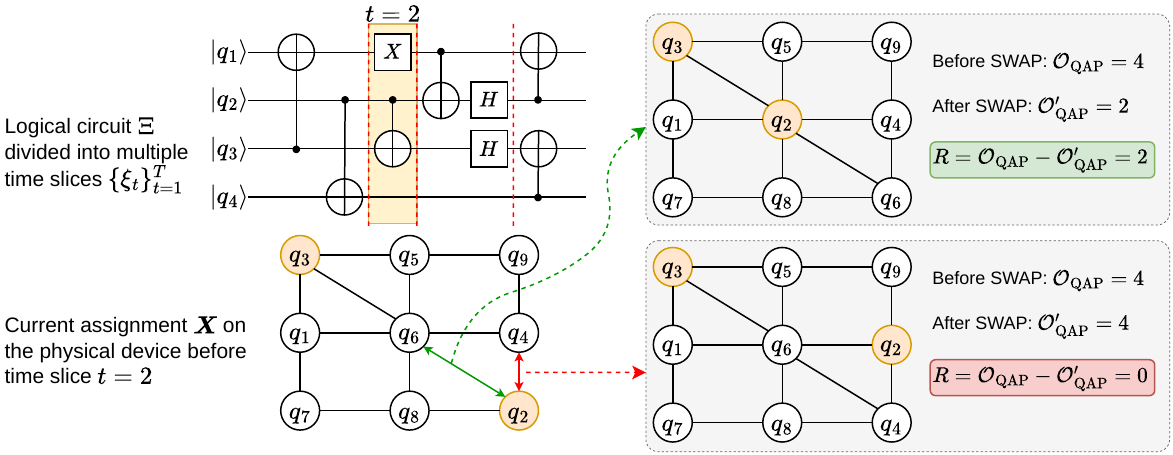}
    \caption{Motivation for integration of QAP objective to the reward function. At time slice $t=2$, we aim to schedule gate $(q_2,q_3)$ by making $q_2$ and $q_3$ adjacent on the device. We compare two actions $\mathrm{SWAP}(q_2,q_6)$ (\textcolor{ForestGreen}{\textbf{green}}) and $\mathrm{SWAP}(q_2,q_4)$ (\textcolor{red}{\textbf{red}}) that result in two reward values. In the former case, the action minimizes the distance between $q_2$ and $q_3$, yielding a positive reward, while the latter case does not change the distance remains unchanged.}
    \label{fig:motivation}
\end{figure}

\subsection{Qubit Routing as a Dynamic QAP Framework}

We adapt the qubit routing problem to the dynamic QAP framework, which can be defined as
\begin{align}
    &\min_{\boldsymbol{X}_1,\dots,\boldsymbol{X}_T}\sum_t^T \mathrm{Tr}(\boldsymbol{F}_t\cdot \boldsymbol{X}_t\cdot \boldsymbol{D}\cdot \boldsymbol{X}_t^\top),\\
    \text{s.t. } &\sum_i \boldsymbol{X}_{t,ij} = 1, \quad\forall j,t,\quad\sum_j \boldsymbol{X}_{t,ij} = 1, \quad\forall i,t\quad \boldsymbol{X}_{t,ij}\in\{0,1\},
\end{align}
where $\boldsymbol{X}_t$ denotes the assignment of logical qubits to physical qubits at the time step $t$, $\boldsymbol{F}_t$ is the flow matrix encoding logical qubit interactions at time $t$, and $\boldsymbol{D}$ is the distance matrix derived from the hardware coupling graph.

A quantum circuit $\Xi$ can be modeled by a Directed Acyclic Graph (DAG), where the nodes correspond to two-qubit gates and the edges represent dependencies between gates.
By partitioning the DAG depthwise according to qubit interaction dependencies, we obtain a sequence of time slices $\{\xi_t\}^T_{t=1}$, where each depth level corresponds to a slice $\xi_t$, consisting of parallel two-qubit gates acting on disjoint qubits (single-qubit gates are omitted).
Formally, we denote the circuit as $\Xi=\{\xi_t\}^T_{t=1}$.

Each slice can be formulated as an individual QAP problem.
At each time step $t$, the permutation $\boldsymbol{X}_t$ determines the physical placement of the logical qubits, and the objective encourages frequently interacting qubits to be mapped to nearby physical nodes (Figure~\ref{fig:motivation}).
Unlike static QAP, the assignments evolve over time, starting from $\boldsymbol{X}_0$, and updated through a sequence of routing operations (e.g. SWAP gates), which transition $\boldsymbol{X}_t$ to $\boldsymbol{X}_{t+1}$.
As a result, the sequential decision process naturally facilitates a reinforcement learning framework, where each action corresponds to a SWAP operation that updates the permutation matrix.

\section{Method}

In this section, we present our reinforcement learning framework for qubit routing. 
We first define the environment, including the observation space, action space, and reward function. 
We then describe the state space encoding procedure within the policy network. 

\subsection{Reinforcement Learning Environment}
Framing the qubit routing problem as reinforcement learning involves defining an observation space and an action space, and designing a proper reward function. 

\vspace{5pt}
\noindent\textbf{Observation space.} The observation space encapsulates the following information:
\begin{enumerate}
    \item The current mapping of logical qubits to physical nodes, or $\boldsymbol{X}_\tau \in \{0,1\}^{N_Q\times N_P}$. 
    \item The current logical interaction graph representing the current time slice $t$, or $\boldsymbol F_\tau\in\mathbb{R}^{N_Q\times N_Q}$.
    \item A series of future time slices of $H$ look-ahead horizon, or $\{\boldsymbol F_{\tau+h}\}^{H}_{h=1}\in\mathbb{R}^{H\times N_Q\times N_Q}$.
    \item The distance matrix of the device graph, or $\boldsymbol D \in\mathbb{R}_{\geq 0}^{N_P\times N_P}$.
\end{enumerate}

\vspace{5pt}
\noindent\textbf{Action space.} 
We partition the circuit $\Xi$ into $T$ time slices and define a maximum $J$ number of steps for each slice $t$ during the rollout.
In other words, we index the environment time step as a tuple $\tau=(t,j)$, where $t\in\{1,\dots,T\}$ denotes the current circuit slice and $j\in\{1,\dots,J\}$ denotes the routing step within that slice.
The action space $\mathcal{A}$ is defined as the set of edges in the physical device, and $\lvert \mathcal{A}\rvert$ is the number of actions.
At any timestep $\tau$, an action $a_\tau=(u,v)\in\mathcal{A}$ indicates a SWAP on the physical device edge $(u,v)$, with $u\neq v; u,v\in\{1,\dots,N_P\}$.
As a result, the observation is updated by applying the selected SWAP action, $\boldsymbol X_{\tau+1} = \boldsymbol X_{\tau}\boldsymbol S_{uv}$,
where $\boldsymbol S_{uv}$ denotes the permutation matrix for the transposition $(u,v)$, i.e., the identity matrix with columns $u$ and $v$ exchanged.
Furthermore, the environment immediately schedules any two-qubit gates whose qubits become adjacent on the device, immediately removing them from the circuit.

In particular, at each routing step $j$ in slice $t$, given the current state $s_\tau=(\boldsymbol X_\tau,\boldsymbol F_\tau,\boldsymbol F_{\tau+1:\tau+H};\boldsymbol D)$, we take an action $a_\tau=(u,v)$ and update the current permutation $\boldsymbol X_\tau$ as $\boldsymbol X_{t,j+1} = \boldsymbol X_{t,j}\boldsymbol S_{uv}$ and the interaction graph $\boldsymbol F_\tau$ as follows.
Let $\mathcal{G}_\tau \subset \{(q_u,q_v):1\leq q_u < q_v\leq N_Q\}$ be the set of executable two-qubit gates after taking action $a_\tau$.
We then define a symmetric gate clearing mask
\begin{align}
    \boldsymbol M_\tau = \sum_{(q_i,q_j)\in\mathcal{G}_\tau} (\boldsymbol e_{q_u}\boldsymbol e_{q_v}^\top + \boldsymbol e_{q_v}\boldsymbol e_{q_u}^\top) \in \{0, 1\}^{N_Q\times N_Q}, \quad \mathrm{diag}(\boldsymbol M_\tau)=0,
\end{align}
where $\boldsymbol e_i\in\mathbb{R}^{N_Q}$ is the $i$-th standard basis vector, and update the interaction graph $\boldsymbol F_\tau$ as
\[
\begin{cases}
    \boldsymbol F_{t,j+1}=\boldsymbol F_{k,j} & \text{if } \mathcal{G}_{t,j}=\emptyset, \\
    \boldsymbol F_{t,j+1}=\boldsymbol F_{t,j} \odot (\mathbf{1} - \boldsymbol M_{t,j}) & \text{otherwise,}
\end{cases}
\]
where $\odot$ denotes the element-wise multiplication (or Hadamard product).

\vspace{5pt}
\noindent\textbf{Reward function.} 
Given the current permutation $\boldsymbol X_{t,j}$, the current interaction graph $\boldsymbol F_{t,j}$ and the new permutation $\boldsymbol X_{t,j+1}$ and the new interaction graph $\boldsymbol F_{t,j+1}$ after taking action $a_{t,j}$, we shape the reward signal based on the QAP objective $\mathcal{O}_\text{QAP}(\boldsymbol X,\boldsymbol F,\boldsymbol D)=\mathrm{Tr}(\boldsymbol F\cdot \boldsymbol X\cdot \boldsymbol D\cdot \boldsymbol X^\top),$ and define the QAP reward term as follows:
\begin{align}
    r_\mathrm{QAP} = \mathcal{O}_\text{QAP}(\boldsymbol X_{t,j},\boldsymbol F_{t,j},\boldsymbol D) - \mathcal{O}_\text{QAP}(\boldsymbol X_{t,j+1},\boldsymbol F_{t,j+1},\boldsymbol D)
\end{align}
The intuition for the subtraction in the QAP term is that $\mathcal{O}_\text{QAP}$ is a minimization objective, so if the flow $\boldsymbol F_{t,j+1}$ after taking the action $a_{t,j}$ yields a lower (better) objective value than $\boldsymbol F_{t,j}$, then the term results in a positive reward.
Since the ultimate objective is to minimize the number of SWAP operations inserted into the circuit, we include a negative constant $\beta$ that encourages the policy to take as few actions as possible.
Lastly, we include a reward for actions that successfully schedule any two-qubit gates proportional to $\lvert \mathcal{G}\rvert$.
The final reward function at time step $\tau=(t,j)$ is
\begin{align}
    r_\tau = \lambda_\text{QAP}\cdot r_\mathrm{QAP} + \lambda_\text{swap} \cdot \beta + \lambda_\text{gate} \cdot \lvert\mathcal{G}_\tau\rvert,
\end{align}
where $\lambda_\text{QAP},\lambda_\text{swap}$ and $\lambda_\text{gate}$ are the balancing coefficients.
To balance the immediate benefit of reducing SWAP overhead in the current time slice against the need to preserve favorable qubit placements for upcoming interactions, we further incorporate a look-ahead mechanism by aggregating future time slices into a decay-weighted effective flow for the QAP term, so that near-term interactions are prioritized while upcoming constraints remain visible.
The combined effective flow matrix is:
\begin{align}
\label{eq:decay-weighted}
    \hat{\boldsymbol{F}}_{t,j} = \sum_{h=0}^{H} \gamma^h \boldsymbol F_{t+h}, \quad 0 < \gamma < 1
\end{align}
This formulation enables the assignment at timestep $\tau$ to be optimized primarily for the current slice $t$, while remaining aware of imminent interactions in future slices.
As a result, the routing strategy favors placements that are immediately effective but do not incur unnecessary overhead when executing upcoming gates, thus balancing short-term optimality with forward-looking consistency.

\begin{figure}
    \centering
    \includegraphics[width=0.9\linewidth]{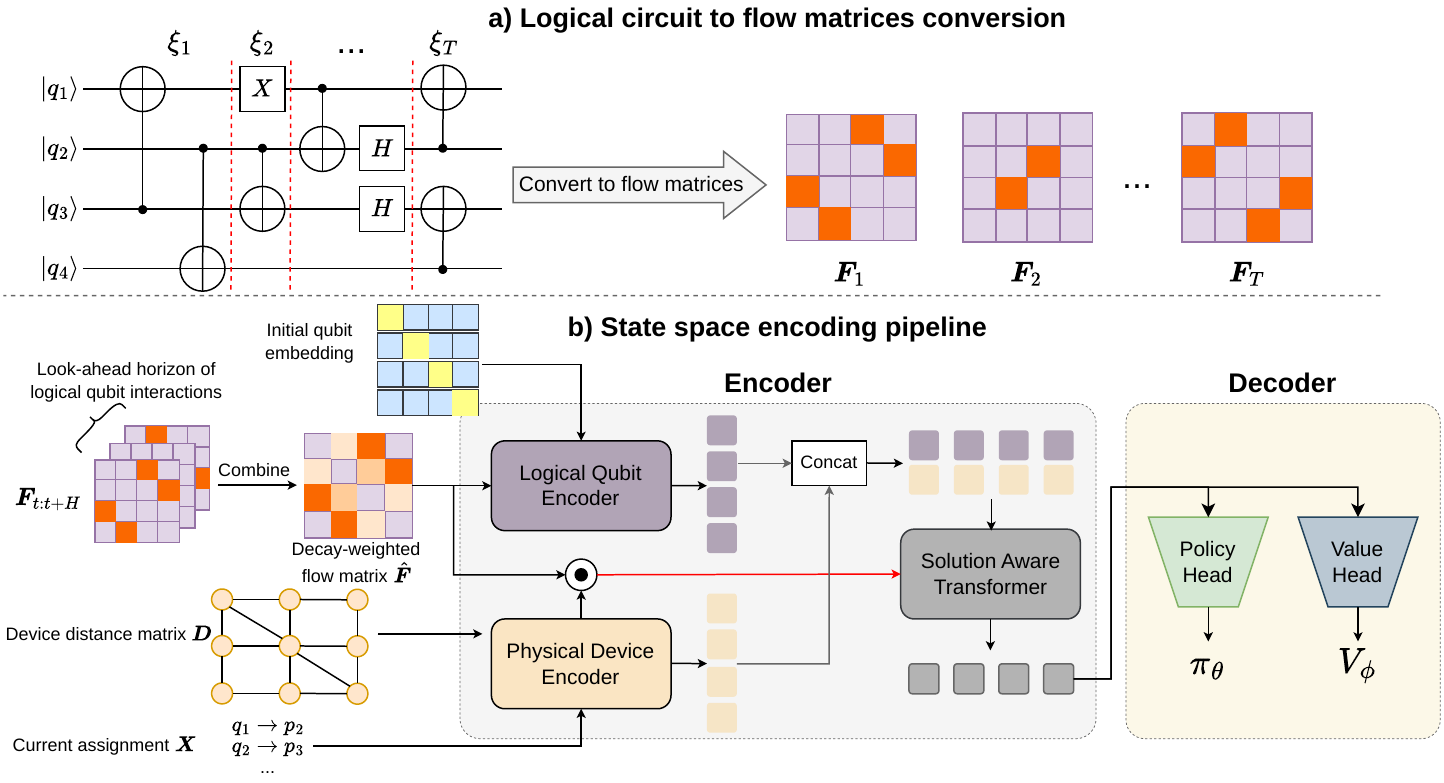}
    \caption{Logical circuit to flow matrices conversion procedure and state space encoding pipeline. (a) A circuit is first divided into a sequence of time slices $\{\xi_t\}_{t=1}^T$, where each slice contain parallelized two-qubit gates (single-qubit gates are omitted). Each slice is then converted to a flow matrix that encode the qubit interactions, yielding $\{\boldsymbol{F}_t\}_{t=1}^T$. (b) The encoding procedure contains a logical qubit encoder that extracts the logical qubit features $\boldsymbol Z_\mathrm{logical}$ via an attention mechanism scaled by a decay-weighted effective flow matrix $\hat{\boldsymbol F}_t$. A physical device encoder extracts the physical node features $\boldsymbol Z_\mathrm{physical}$ governed by the distance matrix $\boldsymbol{D}$ and the current assignment $\boldsymbol{X}_t$. $\boldsymbol Z_\mathrm{logical}$ and $\boldsymbol Z_\mathrm{physical}$ are then concatenated and fed into a solution-aware transformer whose attention matrix is scaled by the QAP objective $\hat{\boldsymbol F}_t \cdot \boldsymbol D$. The resulting context vector $\boldsymbol{Z}_\mathrm{fuse}$ is passed to the policy head $\pi_\theta$ and value head $V_\phi$.}
    \label{fig:placeholder}
\end{figure}

\subsection{State Space Encoding}
The observation space exposes the information required for near‐optimal swap decisions while preserving the Markov property of the underlying decision process.
It provides (i) the current logical to physical assignment, (ii) the current and look‐ahead interaction graphs, and (iii) the hardware topology, as mentioned above.
In this section, we describe how the components in the state space are encoded in the policy network.
There are three modules: a logical qubit encoder and a physical node encoder, both of which are followed by a QAP-aware mixed attention Transformer layer~\cite{vaswani2017attention} that fuses the logical qubit and physical node embeddings, and projects it to the logical space.

\subsubsection{Logical Qubit Encoder}
\label{sec:logical-encoder}

We begin by describing the logical qubit encoder, which extracts representations of logical qubits based on their interaction structure in the circuit.
To encode individual qubits, we assign each qubit a unique identifier via a one-hot encoding.
Each logical qubit $i\in\{1,\dots,N_Q\}$ is assigned a one–hot vector $\boldsymbol E_i\in\mathbb{R}^{N_Q}$; stacking $\{\boldsymbol E_i\}$ yields $\boldsymbol E=\boldsymbol I_{N_Q}$.
The initial qubit features are obtained by a learned projection $\boldsymbol W_E\in\mathbb{R}^{N_Q\times d}$, $
\boldsymbol Z^{(0)} = \boldsymbol E\,\boldsymbol W_E \in \mathbb{R}^{N_Q\times d},$
where $d$ is the hidden dimension.

\noindent\textbf{Flow–aware multi–head attention.}
Let $\boldsymbol F\in\mathbb{R}_{\ge 0}^{N_Q\times N_Q}$ denote the symmetric zero–diagonal flow matrix of an arbitrary time slice.
We perform message passing using a mixed-attention Transformer block whose attention is \emph{scaled} by $\boldsymbol F_\tau$, so that qubits with pending interactions attend to each other.
For layer $\ell$ with $h$ heads and head width $d_k$, we define the query, key and value as
\begin{align}
\label{eq:query-key-value}
\boldsymbol Q_h &= \boldsymbol Z^{(\ell)} \boldsymbol W_{Q,h}, \quad
\boldsymbol K_h = \boldsymbol Z^{(\ell)} \boldsymbol W_{K,h}, \quad
\boldsymbol V_h = \boldsymbol Z^{(\ell)} \boldsymbol W_{V,h}, 
\qquad \boldsymbol W_{(\cdot),h}\in\mathbb{R}^{H\times d_k}.
\end{align}
The head outputs are calculated using the standard Transformer attention $\mathrm{Attn}_h\big(\boldsymbol Z^{(\ell)},\boldsymbol B\big)$, where $\boldsymbol{B}$ is the scaling matrix and $\boldsymbol B\coloneq\boldsymbol F_\tau$.
The multi–head aggregation follows the standard concatenation and then projection procedure. We then apply a residual connection and a position–wise feed–forward network (FFN):
\begin{align}
\label{eq:mha-ffn}
\tilde{\boldsymbol Z}^{(\ell)} = \boldsymbol Z^{(\ell)} + \mathrm{MHA}\left(\mathrm{LN}\big(\boldsymbol Z^{(\ell)}\big);\, \boldsymbol F_\tau\right),\quad \boldsymbol Z^{(\ell+1)} = \tilde{\boldsymbol Z}^{(\ell)} + \mathrm{FFN}\left(\mathrm{LN}\big(\tilde{\boldsymbol Z}^{(\ell)}\big)\right),
\end{align}
where MHA denotes multi-head attention and LN denotes layer normalization~\cite{ba2016layer}.
The final logical embedding is
\begin{equation}
\boldsymbol Z_{\mathrm{logical}} = \mathrm{LN}\big(\boldsymbol Z^{(L)}\big)\in\mathbb{R}^{N_Q\times d}.
\end{equation}
\noindent\textbf{Look–ahead integration.}
With look–ahead enabled, $\boldsymbol F_\tau$ is replaced by a decay–weighted aggregate $\hat {\boldsymbol F}_\tau$ similar to Eq.~\eqref{eq:decay-weighted}, which preserves the same formulation and places greater emphasis on near-term interactions while retaining awareness of future ones.

\subsubsection{Physical Qubit Encoder}
The hardware's physical qubits, referred to as nodes, are specified by the set of physical coordinates $\boldsymbol\varphi\in\mathbb{R}^{N_P\times 2},\boldsymbol\varphi_i\in[0,1]$ and the current logical-to-physical assignment $\boldsymbol X_\tau\in\{0,1\}^{N_Q\times N_P}$. 
The physical qubit encoder first projects the physical coordinates in logical order,
$\boldsymbol\Pi_\tau = \boldsymbol X_\tau\boldsymbol\varphi \in\mathbb{R}^{N_Q\times 2}$,
so that the row $i$ contains the $(x,y)$ location of the logical qubit $i$ on the device at an arbitrary time step.
From $\boldsymbol\Pi_\tau$ and the device coupling graph $G$, we construct a distance matrix based on the shortest-path distances over $G$:
\begin{equation}
\Delta_{\tau ij} = \boldsymbol D\big(\boldsymbol\Pi_{\tau i}, \boldsymbol\Pi_{\tau j}\big),
\qquad
\Delta_\tau \in \mathbb{R}_{\ge 0}^{N_Q \times N_Q},
\end{equation}
with a symmetric row-sum normalization employed,
\begin{equation}
\label{eq:sym-norm}
\bar\Delta_\tau \;=\; \mathrm{diag}(\Delta_\tau \mathbf{1})^{-\tfrac12}\, \Delta_\tau \,\mathrm{diag}(\Delta_\tau \mathbf{1})^{-\tfrac12},
\qquad \bar\Delta_\tau \in \mathbb{R}_{\ge 0}^{N_Q\times N_Q}.
\end{equation}
The initial node features are obtained by projecting coordinates to the hidden vector $\boldsymbol Z^{(0)} =\boldsymbol\Pi\,\boldsymbol W_{\mathrm{proj}} + \mathbf{1}\,\boldsymbol b_{\mathrm{proj}}^\top\in\mathbb{R}^{N_Q\times d}$,
followed by a small stack of \textit{kernel-weighted} residual layers:
\begin{equation}
\label{eq:fcn-prop}
\boldsymbol Z^{(\ell+1)} = \boldsymbol Z^{(\ell)} + \sigma\Big( \bar\Delta_\tau\big(\boldsymbol Z^{(\ell)} \boldsymbol W_\ell + \mathbf{1}\boldsymbol b_\ell^\top\big) \Big),
\end{equation}
where $\boldsymbol W_\ell\in\mathbb{R}^{d\times d}$, $\boldsymbol b_\ell\in\mathbb{R}^{d}$ are learned, and $\sigma$ is an activation function, e.g. ReLU~\cite{agarap2018deep}. After $L$ layers, the final physical node embedding is $\boldsymbol Z_{\mathrm{physical}}=\boldsymbol Z^{(L)}\in\mathbb{R}^{N_Q\times d}$.

\subsubsection{QAP-aware Fusion Layer}
After getting the final logical qubit, $\boldsymbol Z_\mathrm{logical}$, and physical embedding, $\boldsymbol Z_\mathrm{physical}$, we concatenate them and fuse the resulting representation using a feed-forward layer.
\begin{align}
    \boldsymbol Z_\mathrm{fuse} = \mathrm{FFN}([\boldsymbol Z_\mathrm{logical}, \boldsymbol Z_\mathrm{physical}])\in\mathbb{R}^{N_Q\times d}
\end{align}
We then apply a mixed-attention Transformer layer, where the attention $\mathrm{Attn}_h\big(\boldsymbol Z^{(\ell)},\boldsymbol B\big)$ is scaled by the QAP objective matrix formed by the flow matrix $\boldsymbol F_\tau$ and the distance matrix $\boldsymbol D$ , with $\boldsymbol B \coloneq \boldsymbol F_\tau\cdot \boldsymbol D$.
The fused representation $\boldsymbol Z_\mathrm{fuse}$ is then fed into separate policy and value heads, which parameterize a stochastic policy $\pi_\theta(a|s)$ and a value function $V_\phi(s)$, respectively.

\section{Experimental Results}

\subsection{Experiment Setting}

\noindent\textbf{Evaluation metrics.}
In this paper, we focus purely on the qubit routing task by inserting SWAPs into the quantum circuit to make the logical circuit executable on physical hardware.
In the reinforcement learning setting, this corresponds to the number of steps taken in the environment before all gates in the circuit are successfully scheduled.

However, the quantum compilation baselines we compare against may incorporate additional transformations prior to or during routing, such as the use of BRIDGE operations. 
For example, circuit synthesis techniques may introduce additional CNOT gates to effectively relabel qubits or restructure the circuit, thereby reducing routing difficulty at the cost of modifying the original computation~\cite{kremer2024practical} . 
Similarly, BRIDGE operations can replace SWAP sequences with alternative gate constructions that exploit intermediate qubits, trading off gate count and depth in different ways~\cite{cowtan2019qubit}.

Although these techniques can improve overall compilation performance, the distinction between qubit routing and circuit synthesis can become ambiguous. 
To ensure a fair and consistent comparison focused on routing efficiency, we evaluate all methods based on the number of additional two-qubit gates introduced during compilation.
In particular, \textbf{we use the number of added CNOT gates as our primary metric}, as it captures both SWAP and BRIDGE overhead, as well as any auxiliary two-qubit operations introduced by alternative compilation strategies.
In brevity, a SWAP gate can be decomposed into 3 consecutive CNOT gates, and a BRIDGE equals 4 CNOT gates.

\noindent\textbf{Datasets.}
To evaluate our method, we adopt two realistic benchmarks:
\textbf{(i) MQTBench} dataset~\cite{quetschlich2023mqtbench}, or the Munich Quantum Toolkit Benchmark Library, a widely used collection of quantum circuits for quantum compilation.
The dataset includes circuits from a diverse range of algorithms, such as the quantum Fourier transform and the quantum approximate optimization algorithm.
The number of qubits ranges from 2 to 50, with circuit sizes reaching up to 1640 gates.
\textbf{(ii) AgentQ} dataset~\cite{jern2025agent}, a collection of 14000 optimized quantum circuits for 12 different combinatorial optimization problems. 
\textbf{(iii) QUEKO} dataset~\cite{tan2020optimality}, a dataset designed for the qubit mapping task; we focus on the QAOA-MaxCut instances.
We select circuits with 12, 16 and 20 qubits.
To focus exclusively on the routing problem, we represent all two-qubit interactions using CNOT gates and remove single-qubit gates, as they do not affect qubit connectivity or routing requirements.

\vspace{5pt}
\noindent\textbf{Baselines.}
We compare our methods against the following baselines: (i) Greedy method, (ii) Qiskit \texttt{BasicSwap}, (iii) Qiskit \texttt{SabreSwap} with its built-in ``basic'' and ``look-ahead'' heuristics~\cite{li2019tackling,zou2024lightsabre}, (iv) Pytket \texttt{LexiRouting}~\cite{cowtan2019qubit}, and (v) Qiskit \texttt{AIRouting}~\cite{kremer2024practical}.
More details in Appendix~\ref{app:baselines}.

\vspace{5pt}
\noindent\textbf{Physical devices.}
We validate our methods with circuits on the 2D grid and IBM Tokyo-like device topology, with 12, 16, and 20 physical qubits. More details in Appendix~\ref{app:topology}.

\noindent\textbf{Evaluation setup.}
We conduct our experiments in the most challenging setting where $N_Q=N_P$.
We fix an initial mapping for all circuits to ensure a fair comparison.
We adopt the ``trivial'' mapping where logical qubit $i$ is mapped to physical node $i$.
Starting from this mapping, we execute the routing procedure and report the number of inserted CNOT gates.
We further include the results and analyses on individual circuit types and circuit sizes, and on random initial mappings in Appendix~\ref{app:random-mappings}.

\noindent\textbf{Implementation Details}
We use OpenAI \texttt{gymnasium}~\cite{brockman2016openaigym} to implement the RL environment, with \texttt{PyTorch}~\cite{paszke2019pytorch} as the deep learning framework.
We train our policy network on randomly generated circuits to encourage generalization across diverse interaction patterns. 
During training, each circuit is treated as an independent episode in the RL environment, where the agent sequentially selects SWAP operators until all gates in the circuit are scheduled or until a certain number of steps $T_{\max}$ is reached.
We employ a random initial mapping for each episode and train the RL agent with Proximal Policy Optimization~\cite{schulman2017proximal} using \texttt{stable-baselines3}~\cite{raffin2021stable} implementation.
To generate training data, we construct random circuits by sampling two-qubit gates over a set of $N_Q$ logical qubits.
Each gate is then created by uniformly sampling a pair of distinct qubits $(q_u,q_v)$.
This process produces random circuits with varying gate counts and interaction patterns.
During inference, we set $T_{\max} = 1000$.
More details in Appendix~\ref{app:implementation-details}.

\subsection{Results on Qubit Routing}

\begin{table}[]
\centering
\caption{The average number of inserted CNOT gates on MQTBench, AgentQ, and QUEKO datasets with 12, 16 and 20 qubits. The experiments  use the trivial initial mapping on both 2D Grid and IBM Tokyo device topology. Best results are highlighted in \textbf{bold}, and second best in \underline{underline}.}
\resizebox{\linewidth}{!}{
\begin{tabular}{l|cc|cccccc|c}
\toprule
\multirow{2}{*}{Dataset} & Num. of & Num. of & \multirow{2}{*}{Greedy} & \multirow{2}{*}{BasicSwap} & SabreSwap & SabreSwap & \multirow{2}{*}{Pytket} & \multirow{2}{*}{AIRouting} & QAP-Router \\
& qubits & circuits & & & (basic) & (lookahead) & & & (Ours)\\
\midrule
\multicolumn{10}{c}{\textbf{2D Grid}}\\
\midrule
\multirow{3}{*}{MQTBench~\cite{quetschlich2023mqtbench}} & 12 & 152 & 292.32 & 238.48 & 140.17 & 112.97 & 119.47 & \underline{108.09} & \textbf{91.10}\\
&16 & 138 & 722.50 & 510.93 & 273.67 & 222.41 & 226.96 & \underline{216.30} & \textbf{199.84} \\
& 20 & 112 & 1480.45 & 1086.83 & 543.32 & \textbf{421.45} & \underline{427.82} & 432.50 & 428.58\\

\midrule
\multirow{2}{*}{AgentQ~\cite{jern2025agent}}& 12 & 860 & 360.75 & 221.22 & 171.26 & 109.68 & 102.05 & \underline{100.83} & \textbf{70.20}\\
& 16 & 269 & 632.40 & 358.53 & 248.91 & 170.71 & \underline{162.77} & 175.88 & \textbf{133.42} \\

\midrule
\multirow{2}{*}{QUEKO~\cite{tan2020optimality}} & 16 & 150 & 220.16 & 117.26 & 85.32  & \underline{63.52} & 92.30 & 68.88 & \textbf{54.02}\\
& 20 & 150 & 418.96 & 221.20 & 158.16 & \underline{117.28} & 125.38 & 130.36 & \textbf{103.08} \\

\midrule
\multicolumn{10}{c}{\textbf{IBM Tokyo}}\\
\midrule
\multirow{3}{*}{MQTBench~\cite{quetschlich2023mqtbench}} & 12 & 152 & 148.87 & 156.49 & 97.28 & 79.09 & 83.40 & \underline{77.41} & \textbf{77.26}\\
& 16 & 138 & 399.43 & 323.02 & 203.63 & \underline{157.00} & \textbf{153.63} & 162.32 & 166.95 \\
& 20 & 112 & 871.93 & 671.73 & 393.08 & \textbf{300.08} & \underline{308.46} & 338.28 & 323.43 \\

\midrule
\multirow{2}{*}{AgentQ~\cite{jern2025agent}}& 12 & 860 & 172.72 & 106.41 & 104.96 & 89.30 & 78.22 & \underline{67.46} & \textbf{58.05} \\
& 16 & 269 & 363.50 & 214.51 & 171.95 & 129.71 & \underline{122.42} & 134.93 & \textbf{110.19} \\

\midrule
\multirow{2}{*}{QUEKO~\cite{tan2020optimality}} & 16 & 150 & 144.10 & 75.52 & 60.46 & \underline{45.90} & 47.58 & 50.38 & \textbf{42.84} \\
& 20 & 150 & 273.60 & 145.10 & 111.56 & \underline{85.20} & 86.30 & 96.24 & \textbf{75.24} \\
\bottomrule
\end{tabular}
}

\label{tab:num-swaps}
\end{table}


In this section, we report the results on 12-, 16-, and 20-qubit circuits on three datasets: MQTBench~\cite{quetschlich2023mqtbench}, AgentQ~\cite{jern2025agent} and QUEKO~\cite{tan2020optimality}.
We note that comparisons with state-of-the-art routing methods, such as \texttt{SabreSwap} and \texttt{AIRouting}, are not strictly budget-matched, since we can only configure the options exposed through their public APIs. 
Consequently, their internal search and post-processing budgets may differ from ours. 
For instance, even at a lower optimization setting, \texttt{AIRouting} performs 32 routing passes, which is more than an order of magnitude larger than the 3 forward–backward passes used in our method.
Thus, we treat these methods as strong off-the-shelf baselines rather than budget-matched comparisons.

As shown in Table~\ref{tab:num-swaps}, \texttt{QAP-Router} outperforms the state-of-the-art baselines in 11 out of 14 settings on the three datasets across two device topologies. 
On the 2D Grid topology with 16 qubits, it reduces the number of inserted CNOT gates by 7.61\% on MQTBench and 18.4\% on AgentQ.
our method achieves an 11.7\% reduction in inserted CNOT gates compared with \texttt{SabreSwap} and a substantial 21.8\% reduction compared with \texttt{AIRouting}.

We also observe that \texttt{QAP-Router} is less effective in some settings, particularly on 20-qubit MQTBench circuits across both topologies. 
Nevertheless, even in these more challenging cases, \texttt{QAP-Router} remains competitive with the AI-based baseline, Qiskit \texttt{AIRouting}. 
This suggests that while the proposed QAP-guided policy provides strong overall routing performance, further improvements are needed to enhance its robustness on larger and more diverse circuit instances.
The strong performance of \texttt{QAP-Router}, especially on medium-sized circuits, suggests that the proposed QAP-based formulation and solution-aware policy are able to capture globally useful routing structures.
We include more discussion and analysis in Appendix~\ref{app:random-mappings}.


\subsection{Ablation Study on Look-ahead Horizon}
\begin{figure}[hb]
    \centering
    \includegraphics[width=0.75\linewidth]{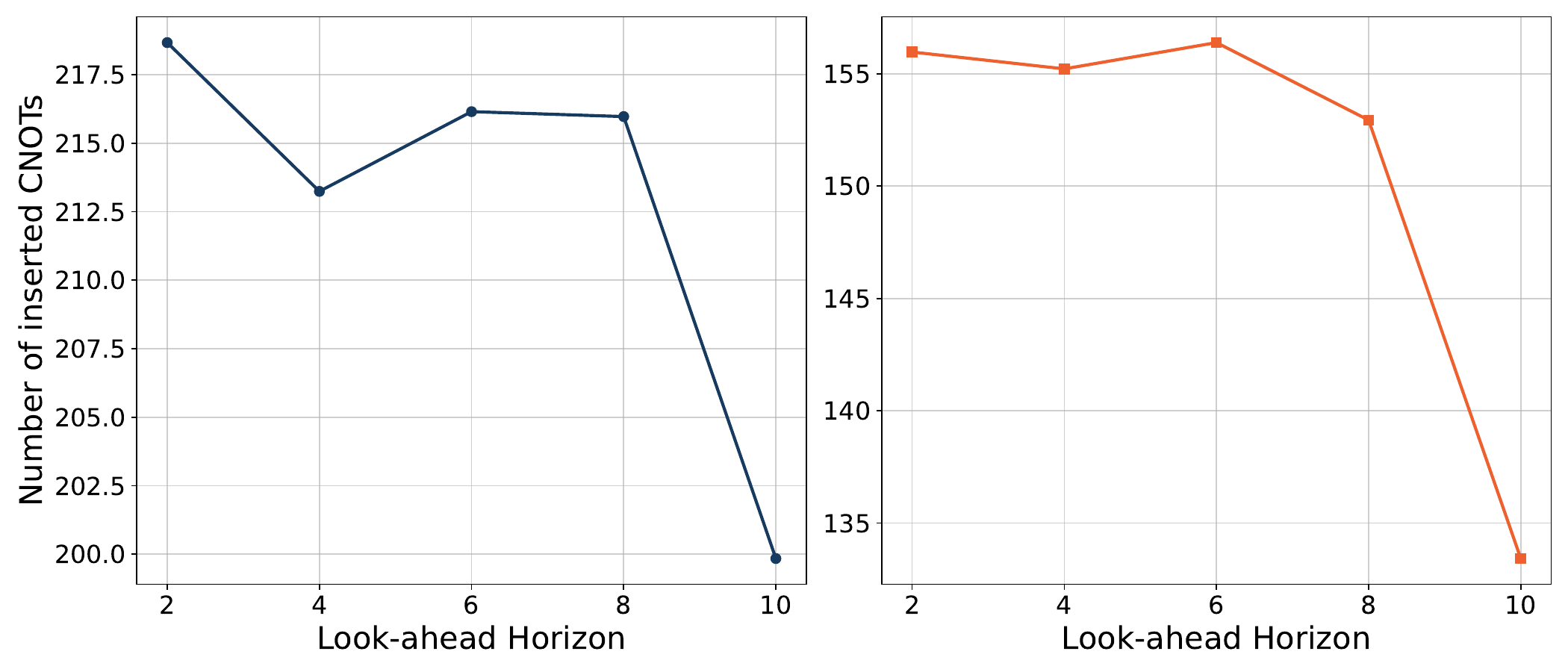}
    \caption{Comparison on the number of inserted CNOT gates across different look-ahead horizons on the 16-qubit 2D Grid device for MQTBench (left) and AgentQ (right) datasets.}
    \label{fig:ablate-lookahead}
\end{figure}

In this ablation study, we examine the effect of the look-ahead horizon length. We consider five different look-ahead horizons, $H \in \{2,4,6,8,10\}$, on the MQTBench and AgentQ datasets in the 16-qubit setting. 
As shown in Figure~\ref{fig:ablate-lookahead}, the performance exhibits a downward trend as $H$ increases, indicating that incorporating a longer look-ahead horizon generally improves routing quality. 
This suggests that exposing the policy to a broader view of upcoming interactions helps it make more globally informed routing decisions, thereby reducing the CNOT overhead in the compiled circuits. 

\section{Conclusion and Future Work}
In this paper, we study the qubit routing problem by proposing a dynamic QAP-based formulation that explicitly captures the interaction-distance coupling between logical circuits and hardware topology.
By integrating this structure into both the reward function and the policy network's encoder, our approach enables efficient routing decisions that balance near-term execution with awareness of future interactions.
Experimental results on realistic circuits demonstrate that our method produces efficient routing strategies with reduced CNOT gate overhead while generalizing across diverse circuit structures.
Such results highlight the benefit of incorporating problem structure into learning-based approaches for quantum compilation.

\noindent\textbf{Future Work.}
We acknowledge several limitations of the proposed method. 
One limitation is that our current framework trains a separate RL policy for each target circuit size, i.e., for each number of qubits. 
While this setting allows the policy to specialize in a fixed action space and device topology, it limits the scalability and reusability of the learned router across circuits of different sizes.
As future work, we aim to develop a more general routing policy that can handle circuits with varying numbers of logical and physical qubits. 
A promising direction is to design size-invariant architectures that operate directly on the circuit interaction graph and hardware coupling graph rather than relying on fixed-size representations. 

\clearpage
\bibliographystyle{plain}
\bibliography{main,ilya-biblio, refs}

@String{Academic = "Academic Press" }

@String{Computer = "{IEEE} Computer" }

@String{Computing = "Computing" }

@STRING{IEEE = {Proc. {IEEE}}}

@STRING{MA = {Meteorological Applications}}

@String{Springer = "Springer-Verlag" }

@article{shaydulin2019network,
  title={Network community detection on small quantum computers},
  author={Shaydulin, Ruslan and Ushijima-Mwesigwa, Hayato and Safro, Ilya and Mniszewski, Susan and Alexeev, Yuri},
  journal={Advanced Quantum Technologies},
  volume={2},
  number={9},
  pages={1900029},
  year={2019},
  publisher={Wiley Online Library},
  doi={10.1002/qute.201900029},
  url={https://dx.doi.org/10.1002/qute.201900029}
}

@article{mitarai2018quantum,
  title={Quantum circuit learning},
  author={Mitarai, K. and Negoro, M. and Kitagawa, M. and Fujii, K.},
  journal={Phys. Rev. A},
  volume={98},
  number={3},
  pages={032309},
  year={2018},
  publisher={APS},
  url={https://journals.aps.org/pra/abstract/10.1103/PhysRevA.98.032309}
}

@article{benedetti2019parameterized,
  title={Parameterized quantum circuits as machine learning models},
  author={Benedetti, Marcello and Lloyd, Erika and Sack, Stefan and Fiorentini, Mattia},
  journal={Quantum Science and Technology},
  volume={4},
  number={4},
  pages={043001},
  year={2019},
  publisher={IOP Publishing}
}

@article{schulman2017proximal,
  title={Proximal policy optimization algorithms},
  author={Schulman, John and Wolski, Filip and Dhariwal, Prafulla and Radford, Alec and Klimov, Oleg},
  journal={arXiv preprint arXiv:1707.06347},
  year={2017}
}

@article{zulehner2018efficient,
  title={An efficient methodology for mapping quantum circuits to the IBM QX architectures},
  author={Zulehner, Alwin and Paler, Alexandru and Wille, Robert},
  journal={IEEE Transactions on Computer-Aided Design of Integrated Circuits and Systems},
  year={2018},
  publisher={IEEE}
}

@article{peruzzo2014variational,
  title={A variational eigenvalue solver on a photonic quantum processor},
  author={Peruzzo, Alberto and McClean, Jarrod and Shadbolt, Peter and Yung, Man-Hong and Zhou, Xiao-Qi and Love, Peter J and Aspuru-Guzik, Al{\'a}n and O’brien, Jeremy L},
  journal={Nature communications},
  volume={5},
  pages={4213},
  year={2014},
  publisher={Nature Publishing Group}
}

@article{biamonte2017quantum,
  title={Quantum machine learning},
  author={Biamonte, Jacob and Wittek, Peter and Pancotti, Nicola and Rebentrost, Patrick and Wiebe, Nathan and Lloyd, Seth},
  journal={Nature},
  volume={549},
  number={7671},
  pages={195},
  year={2017},
  publisher={Nature Publishing Group}
}

@MISC{schulz2017better,
  author = {Schulz, Christian and Tr{\"a}ff, Jesper Larsson},
  title = {Better Process Mapping and Sparse Quadradic Assignment},
  year = {2017},
  journal = {https://arxiv.org/abs/1702.04164}
}

@InProceedings{siraichi2018qubit,
  author    = {Siraichi, Marcos and Dos Santos, Vinicius Fernandes and Collange, Sylv ain and Pereira, Fernando Magno Quint{\~a}o},
  title     = {Qubit Allocation},
  booktitle = {CGO 2018-IEEE/ACM International Symposium on Code Generation and Op timization},
  year      = {2018},
  pages     = {1--12},
}

@article{bach2025efficient,
  title={Efficient Compilation for Shuttling Trapped-Ion Machines via the Position Graph Architectural Abstraction},
  author={Bach, Bao and Safro, Ilya and Younis, Ed},
  journal={arXiv preprint arXiv:2501.12470},
  year={2025}
}

@inproceedings{vaswani2017attention,
  title={Attention is all you need},
  author={Vaswani, Ashish and Shazeer, Noam and Parmar, Niki and Uszkoreit, Jakob and Jones, Llion and Gomez, Aidan N and Kaiser, {\L}ukasz and Polosukhin, Illia},
  booktitle={Advances in neural information processing systems},
  pages={5998--6008},
  year={2017}
}

@article{liu2022leveraging,
  title={Leveraging special-purpose hardware for local search heuristics},
  author={Liu, Xiaoyuan and Ushijima-Mwesigwa, Hayato and Mandal, Avradip and Upadhyay, Sarvagya and Safro, Ilya and Roy, Arnab},
  journal={Computational Optimization and Applications},
  volume={82},
  number={1},
  pages={1--29},
  year={2022},
  publisher={Springer}
}

@article{herman2023quantum,
  title={Quantum computing for finance},
  author={Herman, Dylan and Googin, Cody and Liu, Xiaoyuan and Sun, Yue and Galda, Alexey and Safro, Ilya and Pistoia, Marco and Alexeev, Yuri},
  journal={Nature Reviews Physics},
  volume={5},
  number={8},
  pages={450--465},
  year={2023},
  publisher={Nature Publishing Group UK London}
}

@article{farhi2014quantum,
  title={A quantum approximate optimization algorithm},
  author={Farhi, Edward and Goldstone, Jeffrey and Gutmann, Sam},
  journal={arXiv preprint arXiv:1411.4028},
  year={2014},
  url={https://arxiv.org/abs/arXiv:1411.4028}
}

@Article{abbas2024challenges,
  author    = {Abbas, Amira and Ambainis, Andris and Augustino, Brandon and B{\"a}rtschi, Andreas and Buhrman, Harry and Coffrin, Carleton and Cortiana, Giorgio and Dunjko, Vedran and Egger, Daniel J and Elmegreen, Bruce G and others},
  title     = {Challenges and opportunities in quantum optimization},
  pages     = {1--18},
  journal   = {Nature Reviews Physics},
  publisher = {Nature Publishing Group},
  year      = {2024},
}

@Article{lucas2014ising,
  author    = {Lucas, Andrew},
  title     = {Ising formulations of many NP problems},
  pages     = {74887},
  volume    = {2},
  journal   = {Frontiers in physics},
  publisher = {Frontiers},
  year      = {2014},
}

@article{quetschlich2023mqtbench,
  title={{{MQT Bench}}: Benchmarking Software and Design Automation Tools for Quantum Computing},
  shorttitle = {{MQT Bench}},
  journal = {{Quantum}},
  author={Quetschlich, Nils and Burgholzer, Lukas and Wille, Robert},
  year={2023},
  note={{{MQT Bench}} is available at \url{https://www.cda.cit.tum.de/mqtbench/}},
}

@inproceedings{ito2023algorithmic,
  title={Algorithmic theory of qubit routing},
  author={Ito, Takehiro and Kakimura, Naonori and Kamiyama, Naoyuki and Kobayashi, Yusuke and Okamoto, Yoshio},
  booktitle={Algorithms and Data Structures Symposium},
  pages={533--546},
  year={2023},
  organization={Springer}
}

@article{weinstein2001implementation,
  title={Implementation of the quantum Fourier transform},
  author={Weinstein, Yaakov S and Pravia, MA and Fortunato, EM and Lloyd, Seth and Cory, David G},
  journal={Physical review letters},
  volume={86},
  number={9},
  pages={1889},
  year={2001},
  publisher={APS}
}

@article{herbert2018using,
  title={Using reinforcement learning to find efficient qubit routing policies for deployment in near-term quantum computers},
  author={Herbert, Steven and Sengupta, Akash},
  journal={arXiv preprint arXiv:1812.11619},
  year={2018}
}

@inproceedings{tang2024alpharouter,
  title={Alpharouter: Quantum circuit routing with reinforcement learning and tree search},
  author={Tang, Wei and Duan, Yiheng and Kharkov, Yaroslav and Fakoor, Rasool and Kessler, Eric and Shi, Yunong},
  booktitle={2024 IEEE International Conference on Quantum Computing and Engineering (QCE)},
  volume={1},
  pages={930--940},
  year={2024},
  organization={IEEE}
}

@article{raffin2021stable,
  title={Stable-baselines3: Reliable reinforcement learning implementations},
  author={Raffin, Antonin and Hill, Ashley and Gleave, Adam and Kanervisto, Anssi and Ernestus, Maximilian and Dormann, Noah},
  journal={Journal of machine learning research},
  volume={22},
  number={268},
  pages={1--8},
  year={2021}
}

@article{paszke2019pytorch,
  title={Pytorch: An imperative style, high-performance deep learning library},
  author={Paszke, Adam and Gross, Sam and Massa, Francisco and Lerer, Adam and Bradbury, James and Chanan, Gregory and Killeen, Trevor and Lin, Zeming and Gimelshein, Natalia and Antiga, Luca and others},
  journal={Advances in neural information processing systems},
  volume={32},
  year={2019}
}

@article{yan2024quantum,
  title={Quantum circuit synthesis and compilation optimization: Overview and prospects},
  author={Yan, Ge and Wu, Wenjie and Chen, Yuheng and Pan, Kaisen and Lu, Xudong and Zhou, Zixiang and Wang, Yuhan and Wang, Ruocheng and Yan, Junchi},
  journal={arXiv preprint arXiv:2407.00736},
  year={2024}
}

@inproceedings{sinha2022qubit,
  title={Qubit routing using graph neural network aided Monte Carlo tree search},
  author={Sinha, Animesh and Azad, Utkarsh and Singh, Harjinder},
  booktitle={Proceedings of the AAAI conference on artificial intelligence},
  volume={36},
  number={9},
  pages={9935--9943},
  year={2022}
}

@inproceedings{molavi2022qubit,
  title={Qubit mapping and routing via maxsat},
  author={Molavi, Abtin and Xu, Amanda and Diges, Martin and Pick, Lauren and Tannu, Swamit and Albarghouthi, Aws},
  booktitle={2022 55th IEEE/ACM international symposium on Microarchitecture (MICRO)},
  pages={1078--1091},
  year={2022},
  organization={IEEE}
}

@inproceedings{murali2019noise,
  title={Noise-adaptive compiler mappings for noisy intermediate-scale quantum computers},
  author={Murali, Prakash and Baker, Jonathan M and Javadi-Abhari, Ali and Chong, Frederic T and Martonosi, Margaret},
  booktitle={Proceedings of the twenty-fourth international conference on architectural support for programming languages and operating systems},
  pages={1015--1029},
  year={2019}
}

@article{lee2004generalized,
  title={The generalized quadratic assignment problem},
  author={Lee, Chi-Guhn and Ma, Zhong},
  journal={Research Rep., Dept., Mechanical Industrial Eng., Univ. Toronto, Canada},
  pages={M5S},
  year={2004}
}

@article{lawler1963quadratic,
  title={The quadratic assignment problem},
  author={Lawler, Eugene L},
  journal={Management science},
  volume={9},
  number={4},
  pages={586--599},
  year={1963},
  publisher={INFORMS}
}

@inproceedings{jern2025agent,
  title={Agent-Q: fine-tuning large language models for quantum circuit generation and optimization},
  author={Jern, Linus and Uotila, Valter and Yu, Cong and Zhao, Bo},
  booktitle={2025 IEEE International Conference on Quantum Computing and Engineering (QCE)},
  volume={1},
  pages={1621--1632},
  year={2025},
  organization={IEEE}
}

@article{sivarajah2021t,
  title={t| ket>: a retargetable compiler for NISQ devices},
  author={Sivarajah, Seyon and Dilkes, Silas and Cowtan, Alexander and Simmons, Will and Edgington, Alec and Duncan, Ross},
  journal={Quantum Science \& Technology},
  volume={6},
  number={1},
  pages={014003},
  year={2021},
  publisher={IOP Publishing}
}

@article{karuppasamy2025comprehensive,
  title={A comprehensive review of quantum circuit optimization: Current trends and future directions},
  author={Karuppasamy, Krishnageetha and Puram, Varun and Johnson, Stevens and Thomas, Johnson P},
  journal={Quantum Reports},
  volume={7},
  number={1},
  pages={2},
  year={2025},
  publisher={MDPI}
}

@inproceedings{ushijima2023ising,
  title={An Ising-based Model for Qubit Mapping},
  author={Ushijima Mwesigwa, Hayato and Liu, Xiaoyuan},
  booktitle={Proceedings of the SC'23 Workshops of the International Conference on High Performance Computing, Network, Storage, and Analysis},
  pages={1492--1498},
  year={2023}
}

@article{niu2020hardware,
  title={A hardware-aware heuristic for the qubit mapping problem in the nisq era},
  author={Niu, Siyuan and Suau, Adrien and Staffelbach, Gabriel and Todri-Sanial, Aida},
  journal={IEEE Transactions on Quantum Engineering},
  volume={1},
  pages={1--14},
  year={2020},
  publisher={IEEE}
}

@article{tan2024learning,
  title={Learning solution-aware transformers for efficiently solving quadratic assignment problem},
  author={Tan, Zhentao and Mu, Yadong},
  journal={arXiv preprint arXiv:2406.09899},
  year={2024}
}

@article{tan2020optimality,
  title={Optimality study of existing quantum computing layout synthesis tools},
  author={Tan, Bochen and Cong, Jason},
  journal={IEEE Transactions on Computers},
  volume={70},
  number={9},
  pages={1363--1373},
  year={2020},
  publisher={IEEE}
}

@article{ba2016layer,
  title={Layer normalization},
  author={Ba, Jimmy Lei and Kiros, Jamie Ryan and Hinton, Geoffrey E},
  journal={arXiv preprint arXiv:1607.06450},
  year={2016}
}

@article{portmann2022security,
  title={Security in quantum cryptography},
  author={Portmann, Christopher and Renner, Renato},
  journal={Reviews of Modern Physics},
  volume={94},
  number={2},
  pages={025008},
  year={2022},
  publisher={APS}
}

@article{li2023post,
  title={Post-quantum security: Opportunities and challenges},
  author={Li, Silong and Chen, Yuxiang and Chen, Lin and Liao, Jing and Kuang, Chanchan and Li, Kuanching and Liang, Wei and Xiong, Naixue},
  journal={Sensors},
  volume={23},
  number={21},
  pages={8744},
  year={2023},
  publisher={MDPI}
}

@article{cerezo2022challenges,
  title={Challenges and opportunities in quantum machine learning},
  author={Cerezo, Marco and Verdon, Guillaume and Huang, Hsin-Yuan and Cincio, Lukasz and Coles, Patrick J},
  journal={Nature computational science},
  volume={2},
  number={9},
  pages={567--576},
  year={2022},
  publisher={Nature Publishing Group US New York}
}

@article{zeguendry2023quantum,
  title={Quantum machine learning: A review and case studies},
  author={Zeguendry, Amine and Jarir, Zahi and Quafafou, Mohamed},
  journal={Entropy},
  volume={25},
  number={2},
  pages={287},
  year={2023},
  publisher={MDPI}
}

@article{zhao2026exponential,
  title={Exponential quantum advantage in processing massive classical data},
  author={Zhao, Haimeng and Zlokapa, Alexander and Neven, Hartmut and Babbush, Ryan and Preskill, John and McClean, Jarrod R and Huang, Hsin-Yuan},
  journal={arXiv preprint arXiv:2604.07639},
  year={2026}
}

@misc{fujitsu256qubit,
    title = {Fujitsu and RIKEN develop world-leading 256-qubit superconducting quantum computer},
    author={Fujitsu Limited},
    year = {2025},
    url ={https://info.archives.global.fujitsu/global/about/resources/news/press-releases/2025/0422-01.html},
    note={Date accessed: May 03, 2026. \url{https://info.archives.global.fujitsu/global/about/resources/news/press-releases/2025/0422-01.html}},
}

@misc{brockman2016openaigym,
      title={OpenAI Gym}, 
      author={Greg Brockman and Vicki Cheung and Ludwig Pettersson and Jonas Schneider and John Schulman and Jie Tang and Wojciech Zaremba},
      year={2016},
      eprint={1606.01540},
      archivePrefix={arXiv},
      primaryClass={cs.LG},
      url={https://arxiv.org/abs/1606.01540}, 
}

@incollection{brassard2002quantum,
  title        = {Quantum Amplitude Amplification and Estimation},
  author       = {Brassard, Gilles and H{\o}yer, Peter and Mosca, Michele and Tapp, Alain},
  booktitle    = {Quantum Computation and Information},
  series       = {Contemporary Mathematics},
  volume       = {305},
  pages        = {53--74},
  publisher    = {American Mathematical Society},
  year         = {2002},
  doi          = {10.1090/conm/305/05215},
  eprint       = {quant-ph/0005055},
  archivePrefix= {arXiv}
}

@article{deutsch1992rapid,
  title        = {Rapid Solution of Problems by Quantum Computation},
  author       = {Deutsch, David and Jozsa, Richard},
  journal      = {Proceedings of the Royal Society of London. Series A: Mathematical and Physical Sciences},
  volume       = {439},
  number       = {1907},
  pages        = {553--558},
  year         = {1992},
  doi          = {10.1098/rspa.1992.0167}
}

@incollection{greenberger1989going,
  title        = {Going Beyond {Bell}'s Theorem},
  author       = {Greenberger, Daniel M. and Horne, Michael A. and Zeilinger, Anton},
  booktitle    = {Bell's Theorem, Quantum Theory and Conceptions of the Universe},
  editor       = {Kafatos, Menas},
  pages        = {69--72},
  publisher    = {Kluwer Academic Publishers},
  address      = {Dordrecht},
  year         = {1989},
  doi          = {10.1007/978-94-017-0849-4_10}
}

@incollection{hein2006entanglement,
  title        = {Entanglement in Graph States and Its Applications},
  author       = {Hein, Marc and D{\"u}r, Wolfgang and Eisert, Jens and Raussendorf, Robert and Van den Nest, Maarten and Briegel, Hans J.},
  booktitle    = {Quantum Computers, Algorithms and Chaos},
  editor       = {Casati, Giulio and Shepelyansky, Dima L. and Zoller, Peter and Benenti, Giuliano},
  series       = {Proceedings of the International School of Physics ``Enrico Fermi''},
  volume       = {162},
  pages        = {115--218},
  publisher    = {IOS Press},
  year         = {2006},
  doi          = {10.3254/978-1-61499-018-5-115},
  eprint       = {quant-ph/0602096},
  archivePrefix= {arXiv}
}

@article{raussendorf2001oneway,
  title        = {A One-Way Quantum Computer},
  author       = {Raussendorf, Robert and Briegel, Hans J.},
  journal      = {Physical Review Letters},
  volume       = {86},
  number       = {22},
  pages        = {5188--5191},
  year         = {2001},
  doi          = {10.1103/PhysRevLett.86.5188}
}

@article{farhi2014qaoa,
  title        = {A Quantum Approximate Optimization Algorithm},
  author       = {Farhi, Edward and Goldstone, Jeffrey and Gutmann, Sam},
  year         = {2014},
  eprint       = {1411.4028},
  archivePrefix= {arXiv},
  primaryClass = {quant-ph}
}

@article{cleve1998quantum,
  title        = {Quantum Algorithms Revisited},
  author       = {Cleve, Richard and Ekert, Artur and Macchiavello, Chiara and Mosca, Michele},
  journal      = {Proceedings of the Royal Society of London. Series A: Mathematical, Physical and Engineering Sciences},
  volume       = {454},
  number       = {1969},
  pages        = {339--354},
  year         = {1998},
  doi          = {10.1098/rspa.1998.0164},
  eprint       = {quant-ph/9708016},
  archivePrefix= {arXiv}
}

@article{dur2000three,
  title        = {Three Qubits Can Be Entangled in Two Inequivalent Ways},
  author       = {D{\"u}r, Wolfgang and Vidal, Guifr{\'e} and Cirac, J. Ignacio},
  journal      = {Physical Review A},
  volume       = {62},
  number       = {6},
  pages        = {062314},
  year         = {2000},
  doi          = {10.1103/PhysRevA.62.062314},
  eprint       = {quant-ph/0005115},
  archivePrefix= {arXiv}
}

@misc{qiskitrealamplitudes,
  title        = {{RealAmplitudes}: {Qiskit} Circuit Library Documentation},
  author       = {{IBM Quantum}},
  year         = {2026},
  howpublished = {\url{https://quantum.cloud.ibm.com/docs/api/qiskit/qiskit.circuit.library.RealAmplitudes}},
  note         = {Accessed: 2026-05-05}
}

@misc{qiskitefficientsu2,
  title        = {{EfficientSU2}: {Qiskit} Circuit Library Documentation},
  author       = {{IBM Quantum}},
  year         = {2026},
  howpublished = {\url{https://quantum.cloud.ibm.com/docs/api/qiskit/qiskit.circuit.library.EfficientSU2}},
  note         = {Accessed: 2026-05-05}
}

@misc{qiskittwolocal,
  title        = {{TwoLocal}: {Qiskit} Circuit Library Documentation},
  author       = {{IBM Quantum}},
  year         = {2026},
  howpublished = {\url{https://quantum.cloud.ibm.com/docs/api/qiskit/qiskit.circuit.library.TwoLocal}},
  note         = {Accessed: 2026-05-05}
}

@misc{qiskitbasicswap,
  title        = {{BasicSwap}: {Qiskit} Transpiler Pass Documentation},
  author       = {{IBM Quantum}},
  year         = {2026},
  howpublished = {\url{https://quantum.cloud.ibm.com/docs/api/qiskit/qiskit.transpiler.passes.BasicSwap}},
  note         = {Accessed: 2026-05-05}
}

@inproceedings{sudheer2012optimization,
  title={Optimization of the hop-byte metric for effective topology aware mapping},
  author={Sudheer, C Devi and Srinivasan, Ashok},
  booktitle={2012 19th International Conference on High Performance Computing},
  pages={1--9},
  year={2012},
  organization={IEEE}
}

@inproceedings{hoefler2011generic,
  title={Generic topology mapping strategies for large-scale parallel architectures},
  author={Hoefler, Torsten and Snir, Marc},
  booktitle={Proceedings of the international conference on Supercomputing},
  pages={75--84},
  year={2011}
}

@article{agarap2018deep,
  title={Deep learning using rectified linear units (relu)},
  author={Agarap, Abien Fred},
  journal={arXiv preprint arXiv:1803.08375},
  year={2018}
}

@article{bagga2023solving,
  title={Solving the quadratic assignment problem using deep reinforcement learning},
  author={Bagga, Puneet S and Delarue, Arthur},
  journal={arXiv preprint arXiv:2310.01604},
  year={2023}
}

@misc{bengio2020co,
      title={Machine Learning for Combinatorial Optimization: a Methodological Tour d'Horizon}, 
      author={Yoshua Bengio and Andrea Lodi and Antoine Prouvost},
      year={2020},
      eprint={1811.06128},
      archivePrefix={arXiv},
      primaryClass={cs.LG},
      url={https://arxiv.org/abs/1811.06128}, 
}

@inproceedings{nowak2018revised,
  title={Revised note on learning quadratic assignment with graph neural networks},
  author={Nowak, Alex and Villar, Soledad and Bandeira, Afonso S and Bruna, Joan},
  booktitle={2018 IEEE Data Science Workshop (DSW)},
  pages={1--5},
  year={2018},
  organization={IEEE}
}

@article{wang2021neural,
  title={Neural graph matching network: Learning lawler’s quadratic assignment problem with extension to hypergraph and multiple-graph matching},
  author={Wang, Runzhong and Yan, Junchi and Yang, Xiaokang},
  journal={IEEE Transactions on Pattern Analysis and Machine Intelligence},
  volume={44},
  number={9},
  pages={5261--5279},
  year={2021},
  publisher={IEEE}
}

@article{liu2020revocable,
  title={Revocable deep reinforcement learning with affinity regularization for outlier-robust graph matching},
  author={Liu, Chang and Jiang, Zetian and Wang, Runzhong and Yan, Junchi and Huang, Lingxiao and Lu, Pinyan},
  journal={arXiv preprint arXiv:2012.08950},
  year={2020}
}

@article{molavi2026generating,
  title={Generating Compilers for Qubit Mapping and Routing},
  author={Molavi, Abtin and Xu, Amanda and Cecchetti, Ethan and Tannu, Swamit and Albarghouthi, Aws},
  journal={Proceedings of the ACM on Programming Languages},
  volume={10},
  number={POPL},
  pages={2265--2294},
  year={2026},
  publisher={ACM New York, NY, USA}
}

@article{javadi2024quantum,
  title={Quantum computing with Qiskit},
  author={Javadi-Abhari, Ali and Treinish, Matthew and Krsulich, Kevin and Wood, Christopher J and Lishman, Jake and Gacon, Julien and Martiel, Simon and Nation, Paul D and Bishop, Lev S and Cross, Andrew W and others},
  journal={arXiv preprint arXiv:2405.08810},
  year={2024}
}

@article{pozzi2022using,
  title={Using reinforcement learning to perform qubit routing in quantum compilers},
  author={Pozzi, Matteo G and Herbert, Steven J and Sengupta, Akash and Mullins, Robert D},
  journal={ACM Transactions on Quantum Computing},
  volume={3},
  number={2},
  pages={1--25},
  year={2022},
  publisher={ACM New York, NY}
}

@inproceedings{li2019tackling,
  title={Tackling the qubit mapping problem for NISQ-era quantum devices},
  author={Li, Gushu and Ding, Yufei and Xie, Yuan},
  booktitle={Proceedings of the twenty-fourth international conference on architectural support for programming languages and operating systems},
  pages={1001--1014},
  year={2019}
}

@incollection{maronese2022quantum,
  title={Quantum compiling},
  author={Maronese, Marco and Moro, Lorenzo and Rocutto, Lorenzo and Prati, Enrico},
  booktitle={Quantum Computing Environments},
  pages={39--74},
  year={2022},
  publisher={Springer}
}

@article{smith2020open,
  title={An open-source, industrial-strength optimizing compiler for quantum programs},
  author={Smith, Robert S and Peterson, Eric C and Skilbeck, Mark G and Davis, Erik J},
  journal={Quantum Science \& Technology},
  volume={5},
  number={4},
  pages={044001},
  year={2020},
  publisher={IOP Publishing}
}

@article{cowtan2019qubit,
  title={On the qubit routing problem},
  author={Cowtan, Alexander and Dilkes, Silas and Duncan, Ross and Krajenbrink, Alexandre and Simmons, Will and Sivarajah, Seyon},
  journal={arXiv preprint arXiv:1902.08091},
  year={2019}
}

@article{yolcu2019learning,
  title={Learning local search heuristics for boolean satisfiability},
  author={Yolcu, Emre and P{\'o}czos, Barnab{\'a}s},
  journal={Advances in Neural Information Processing Systems},
  volume={32},
  year={2019}
}

@article{kremer2024practical,
  title={Practical and efficient quantum circuit synthesis and transpiling with reinforcement learning},
  author={Kremer, David and Villar, Victor and Paik, Hanhee and Duran, Ivan and Faro, Ismael and Cruz-Benito, Juan},
  journal={arXiv preprint arXiv:2405.13196},
  year={2024}
}

@article{kim2023evidence,
  title={Evidence for the utility of quantum computing before fault tolerance},
  author={Kim, Youngseok and Eddins, Andrew and Anand, Sajant and Wei, Ken Xuan and Van Den Berg, Ewout and Rosenblatt, Sami and Nayfeh, Hasan and Wu, Yantao and Zaletel, Michael and Temme, Kristan and others},
  journal={Nature},
  volume={618},
  number={7965},
  pages={500--505},
  year={2023},
  publisher={Nature Publishing Group UK London}
}

@article{zou2024lightsabre,
  title={LightSABRE: A lightweight and enhanced SABRE algorithm},
  author={Zou, Henry and Treinish, Matthew and Hartman, Kevin and Ivrii, Alexander and Lishman, Jake},
  journal={arXiv preprint arXiv:2409.08368},
  year={2024}
}

@inproceedings{huang2022reinforcement,
  title={Reinforcement learning and dear framework for solving the qubit mapping problem},
  author={Huang, Ching-Yao and Lien, Chi-Hsiang and Mak, Wai-Kei},
  booktitle={Proceedings of the 41st IEEE/ACM international conference on computer-aided design},
  pages={1--9},
  year={2022}
}

\appendix



\section{Related Work}

Qubit mapping and routing are of central importance in quantum compilation. These problems are often defined in terms of boolean satisfiability and often require circuit transformations to achieve a good enough mapping/routing~\cite{siraichi2018qubit}. Different quantum compilers use different types of qubit mapping and routing algorithms to achieve a low overhead. Some examples include heuristic search~\cite{yolcu2019learning, javadi2024quantum}, graph guided search algorithms~\cite{smith2020open}, quadratic binary unconstrained optimization (QUBO)~\cite{liu2022leveraging} and template matching~\cite{molavi2026generating}. Recently, some approaches have explored applying RL to this class of problems by either reducing it to sequential decision making~\cite{kremer2024practical} or by modeling it as sequence-to-sequence task~\cite{huang2022reinforcement}. 

Our method is inspired by several works in the domain of RL algorithms applied to CO problems~\cite{bengio2020co}. In this work, we leverage graph neural networks to solve QAP problem. The use of Graph Neural Networks is similar to existing work~\cite{wang2021neural,nowak2018revised,liu2020revocable} which use these networks to solve QAP by graph matching. However, our approach does not incur the expensive $O(n^4)$ graph matching cost. Similarly, RL based QAP solvers~\cite{bagga2023solving, tan2024learning} focus on solving a static QAP problem. In contrast, in this work we tackle the harder problem of a \emph{dynamic} QAP. Taken in context of qubit routing, our method is novel in itself. However, we believe that the algorithm presented here has the potential to be a general dynamic QAP solver as well.

\section{Physical Device Topology}
\label{app:topology}
\begin{figure}
    \centering
    \includegraphics[width=0.85\linewidth]{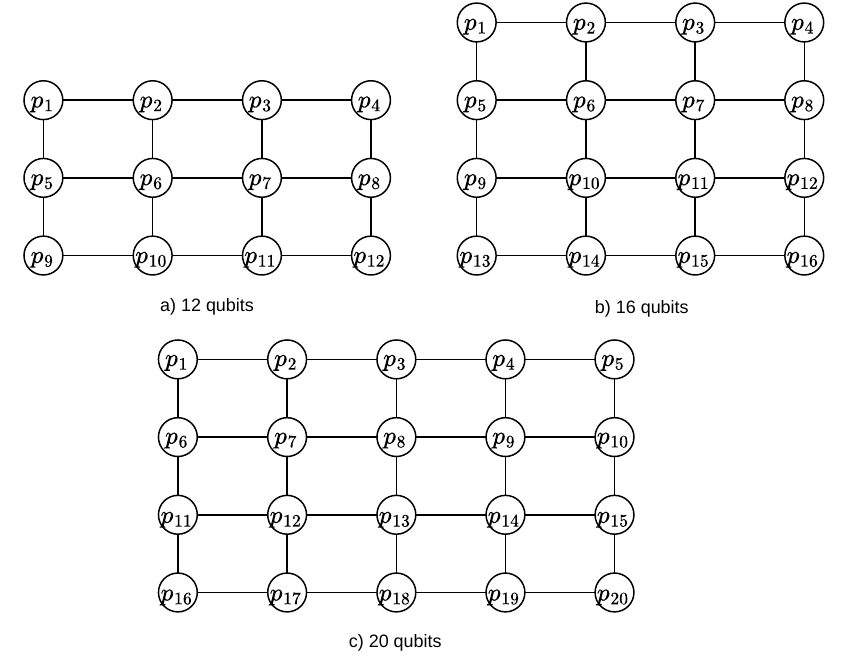}
    \caption{2D Grid device topology for 12, 16 and 20 qubits.}
    \label{fig:grid}
\end{figure}

\noindent\textbf{2D Grid.}
We consider a regular 2D grid topology in which qubits are arranged on a lattice (Figure~\ref{fig:grid}), and each qubit is connected to its nearest neighbors (up to four). 
This corresponds to rectangular subgrids that provide uniform connectivity and serve as a simplified abstraction of hardware constraints.

\begin{figure}
    \centering
    \includegraphics[width=0.85\linewidth]{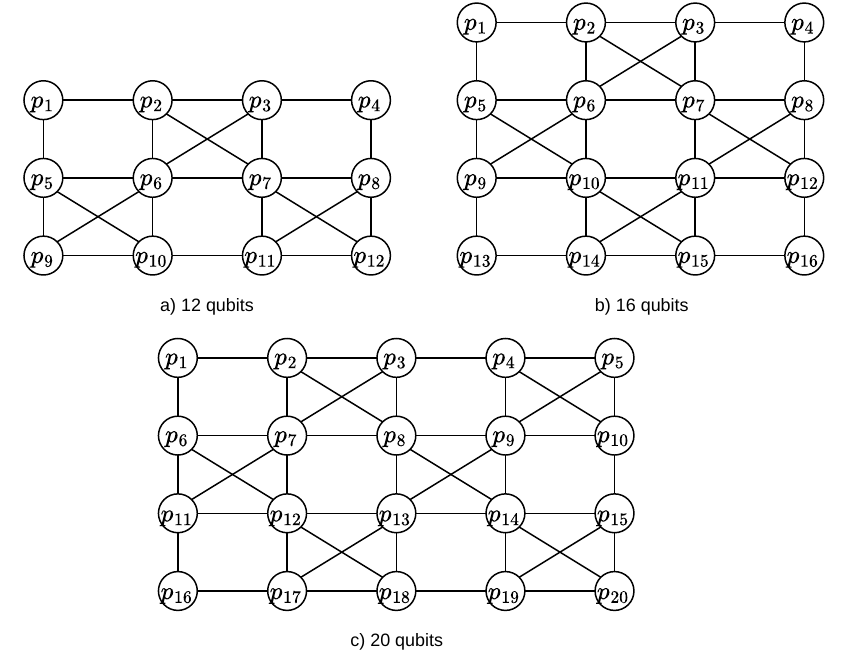}
    \caption{IBM Tokyo-like device topology for 12, 16 and 20 qubits.}
    \label{fig:tokyo}
\end{figure}

\noindent\textbf{IBM Tokyo.}
We also evaluate our method on hardware-inspired topologies based on IBM devices, specifically 12- and 16-qubit subgraphs extracted from larger architectures such as IBM Tokyo (originally with 20 qubits). 
These topologies feature irregular connectivity with varying node degrees  (Figure~\ref{fig:tokyo}), reflecting the realistic constraints of superconducting quantum processors and posing more challenging routing scenarios compared to regular grids.
\begin{itemize}
    \item \textbf{12-qubit} additional edges
    \begin{align*}
    E_{\text{extra}} = \{(1,6), (2,5), (4,9), (5,8), (6,11), (7,10)\}.
    \end{align*}
    \item \textbf{16-qubit} additional edges
    \begin{align*}
    E_{\text{extra}} = \{(1,6), (2,5), (4,9), (5,8), (6,11), (7,10), (9,14), (10, 13)\}.
    \end{align*}
    \item \textbf{20-qubit} additional edges
    \begin{align*}
        E_{\text{extra}} = &\{(1,7), (2,6), (3,9), (4,8), (5,11), (6,10),\\ &(7,13), (8,12), (11,17), (12,16), (13,19), (14,18)\}.
    \end{align*}

\end{itemize}

\section{Baseline Descriptions}\label{app:baselines}
In this appendix section, we briefly describe the baselines used in the empirical study.
Let $\pi$ be the current logical-to-physical qubit mapping, $F_\mathrm{front}$ be the current front layer of executable two-qubit gates, and $\mathcal{G}_\mathrm{future}$ be a set of future gates used for look-ahead scoring.

\noindent\textbf{Pytket \texttt{LexiRouting}.}
The \texttt{LexiRouting} baseline uses the routing procedure implemented in \textsc{tket}/Pytket. 
In our experiments, this corresponds to applying Pytket's routing framework with \texttt{LexiRouteRoutingMethod}.
The method updates the mapping by inserting SWAP operations and may also use architecture-aware replacements such as BRIDGE operations when applicable. 
It includes a look-ahead parameter that controls how many upcoming two-qubit gates are considered when selecting routing modifications~\cite{cowtan2019qubit}.
We set the look-ahead horizon to 10.

\noindent\textbf{Greedy.}
The greedy baseline is Pytket \texttt{LexiRouting} with the look-ahead horizon set to 0.

\noindent\textbf{Qiskit \texttt{BasicSwap}.}
\texttt{BasicSwap} is Qiskit's minimum-effort SWAP insertion pass. 
It traverses the circuit and checks whether each two-qubit operation is compatible with the target coupling map under the current layout. 
When a two-qubit operation acts on non-adjacent physical qubits, \texttt{BasicSwap} inserts one or more SWAP gates before the operation to make it executable. 
In Qiskit's implementation, these SWAPs are inserted along a shortest undirected path between the two physical qubits involved in the operation. 
Since this method does not perform a global search or look-ahead optimization, it is typically simple and deterministic, but may introduce more SWAPs than more sophisticated routing methods~\cite{qiskitbasicswap}.

\noindent\textbf{Qiskit \texttt{SabreSwap} with ``basic'' heuristic.}
\texttt{SabreSwap} is Qiskit's implementation of the SABRE routing algorithm. SABRE maintains a front layer of two-qubit gates and searches over candidate SWAPs in the neighborhood of qubits involved in the front layer. With the ``basic'' heuristic, each candidate SWAP is scored by the sum of physical distances between the qubits of the front-layer gates after applying that SWAP:
\begin{align}
    \mathrm{SabreSwap}_{\mathrm{basic}}
    =
    \sum_{(q_i,q_j)\in F_\mathrm{front}}
    \boldsymbol D\big(\pi(q_i), \pi(q_j)\big).
\end{align}
The SWAP with the lowest score is selected, the layout is updated, and the process repeats until all gates are routed. This heuristic focuses only on the current front layer and therefore provides a direct comparison against methods that incorporate future interactions~\cite{li2019tackling}.

\noindent\textbf{Qiskit \texttt{SabreSwap} with ``look-ahead'' heuristic.}
The ``look-ahead'' variant of \texttt{SabreSwap} extends the basic SABRE cost by incorporating an additional set of upcoming gates. Specifically, it combines the front-layer cost with a weighted cost over an extended set $\mathcal{G}_\mathrm{future}$ of future gates:
\begin{align}
    \mathrm{SabreSwap}_{\mathrm{lookahead}}
    =
    \frac{1}{|F_\mathrm{front}|}
    &\sum_{(q_i,q_j)\in F_\mathrm{front}}
    \boldsymbol D\big(\pi(q_i), \pi(q_j)\big) \\
    &+
    \omega\cdot
    \frac{1}{|\mathcal{G}_\mathrm{future}|}
    \sum_{(q_i,q_j)\in \mathcal{G}_\mathrm{future}}
    \boldsymbol D\big(\pi(q_i), \pi(q_j)\big),
\end{align}
where $\omega$ is a weighting coefficient that gives lower priority to future gates than to immediately executable gates. Compared with the ``basic'' heuristic, this variant is less myopic because it attempts to choose SWAPs that are beneficial for both the current front layer and near-future interactions~\cite{li2019tackling}. We set the number of trials to 1, which is equivalent to 3 passes.

\noindent\textbf{Qiskit \texttt{AIRouting}.}
\texttt{AIRouting} is an AI-powered routing pass from the \texttt{qiskit-ibm-transpiler} package. Unlike purely hand-designed heuristics such as \texttt{BasicSwap} and \texttt{SabreSwap}, \texttt{AIRouting} uses reinforcement-learning-based models for layout selection and circuit routing. The pass acts as both a layout stage and a routing stage: depending on the selected layout mode, it may keep, improve, or optimize the initial layout before routing the circuit. We use \texttt{AIRouting} as an off-the-shelf learned routing baseline and compare its routed circuits against heuristic baselines and our proposed method~\cite{kremer2024practical}. We set its optimization level to 1, which is equivalent to doing 32 passes.

\section{Additional Implementation Details}
\label{app:implementation-details}
We use OpenAI \texttt{gymnasium}~\cite{brockman2016openaigym} to implement the RL environment, with the integration of \texttt{PyTorch}~\cite{paszke2019pytorch} as the deep learning framework.
We train our policy network on randomly generated circuits to encourage generalization across diverse interaction patterns. 
During training, each circuit is treated as an independent episode in the RL environment, where the agent sequentially selects SWAP operators until all gates in the circuit are scheduled.
We employ a random initial mapping for each episode to enhance robustness and encourage the policy to generalize across arbitrary mappings.

We train the RL policy network with Proximal Policy Optimization (PPO)~\cite{schulman2017proximal} using \texttt{stable-baselines3}~\cite{raffin2021stable} implementation.
We adopt the default decoder architecture from \texttt{stable-baselines3}, a two-layer feed forward network, and set the hidden dimension to 256.
We use a learning rate of 0.0003, a batch size of 1024, and a total number of training timesteps of 10M on 12 qubits, 20M on 16, and 50M on 20.
Although these budgets define the maximum number of environment interactions, we empirically observe that the learned policy generally converges after approximately one third of the allocated timesteps.
The remaining hyper-parameters for PPO follow by the library default settings.

To choose the best set of hyper-parameters $(\lambda_\mathrm{QAP}, \lambda_\mathrm{swap},\lambda_\mathrm{gate})$, we perform a grid search over the values $\{0.5, 1.0, 2.0\}$, yielding a total of 27 runs per setting.
Empirically, we observe the best combination to be $(\lambda_\mathrm{QAP}, \lambda_\mathrm{swap},\lambda_\mathrm{gate})=(1.0,2.0,2.0)$.
All experiments were conducted on a single 141GB NVIDIA H200 GPU.

\noindent\textbf{Training Data Generation.} To generate training data, we construct random circuits by sampling two-qubit gates over a set of $N_Q$ logical qubits.
Specifically, we first sample the total number of two-qubit gates uniformly at random between $\kappa_\text{low}\cdot N_Q$ and $\kappa_\text{high}\cdot N_Q$, where $\kappa_\text{low},\kappa_\text{high}$ denote low and high thresholds.
Empirically, we set $(\kappa_\text{low},\kappa_\text{high})=(8,16)$.
Each gate is then created by uniformly sampling a pair of distinct qubits $(q_u,q_v)$.
This process produces random circuits with varying gate counts and interaction patterns.

\noindent\textbf{Post-processing.}
During evaluation, we employ a lightweight post-processing step following the conventional bidirectional routing strategy used in SABRE~\cite{li2019tackling}. 
The purpose of this step is to refine the final routed circuit without introducing an expensive search procedure.
Specifically, for each circuit, we perform three routing passes in a forward--backward--forward order. 
In the first forward pass, the circuit is routed from the initial mapping $\boldsymbol{X}_0$ order to obtain an initial feasible layout $\boldsymbol{X}'_0$ and routed circuit.
The second pass routes the circuit in the reversed order of gates using the final layout from the first pass $\boldsymbol{X}'_0$ as its initial layout, which helps propagate information from later circuit interactions back toward the beginning of the circuit.
We also record the final layout of the backward pass as $\boldsymbol{X}''_0$.
Finally, a third forward pass routes the original circuit again using the layout information obtained from the backward pass $\boldsymbol{X}''_0$. 

This bidirectional procedure allows the router to account for both early and late two-qubit interactions. 
Compared with applying a single forward pass, the forward--backward--forward refinement can reduce unnecessary SWAP insertions caused by myopic layout choices, while adding only a small constant-factor overhead during evaluation. 
All reported results use the routed circuit produced by the final forward pass.

\section{More Results on Random Initial Mappings}\label{app:random-mappings}
To further evaluate the robustness of our method, we test \texttt{QAP-Router} under random initial mappings. 
Specifically, we generate a fixed set of 10 random initial mappings and report the averaged results in Table~\ref{tab:random-mappings} on two datasets, MQTBench and QUEKO, using the more challenging 2D Grid device topology. 
As shown in Table~\ref{tab:random-mappings}, \texttt{QAP-Router} is robust to variations in the initial mapping, achieving the best performance in 4 out of 5 settings. 
Although our method struggles on 20-qubit MQTBench circuits under the trivial initial mapping, as reported in Table~\ref{tab:num-swaps}, it achieves the second-best result under random initial mappings, trailing \texttt{SabreSwap} by only about 12 CNOT gates, equivalent to approximately 4 SWAP gates. 
These results suggest that \texttt{QAP-Router} is not overly dependent on a favorable initial layout and can generalize effectively across different starting mappings.

\begin{table}[]
\centering
\caption{The average number of inserted CNOT gates on MQTBench, AgentQ, and QUEKO datasets with 12, 16 and 20 qubits. The experiments use a fixed set of 10 random initial mappings on 2D Grid device topology. Best results are highlighted in \textbf{bold}, and second best in \underline{underline}.}
\resizebox{\linewidth}{!}{
\begin{tabular}{l|cc|ccccc|c}
\toprule
\multirow{2}{*}{Dataset} & Num. of & Num. of  & \multirow{2}{*}{BasicSwap} & SabreSwap & SabreSwap & \multirow{2}{*}{Pytket} & \multirow{2}{*}{AIRouting} & QAP-Router \\
 & qubits & circuits & & (basic) & (lookahead) & & & (Ours)\\
\midrule
\multirow{3}{*}{MQTBench} & 12 & 152 & 226.48 $\pm$ 8.92 & 145.61 $\pm$ 1.95 & 113.33 $\pm$ 3.12 & 110.87 $\pm$ 5.38 & \underline{109.74 $\pm$ 2.08} & \textbf{101.64 $\pm$ 3.54} \\
&16 & 138 & 519.00 $\pm$ 27.51 & 301.88 $\pm$ 4.17 & \underline{227.00 $\pm$ 3.93} & 233.88 $\pm$ 6.50 & 234.41 $\pm$ 3.87 & \textbf{215.40 $\pm$ 4.53} \\
& 20 & 112 & 1046.47 $\pm$ 45.18 & 572.98 $\pm$ 11.27 & \textbf{424.33 $\pm$ 9.58} & 441.70 $\pm$ 10.14 & 450.46 $\pm$ 10.45 & \underline{436.29 $\pm$ 7.22} \\

\midrule
\multirow{2}{*}{QUEKO} & 16 & 150 & 122.07 $\pm$ 1.91 & 89.52 $\pm$ 2.21 & \underline{65.93 $\pm$ 1.50} & 68.79 $\pm$ 1.46 & 71.72 $\pm$ 1.66 & \textbf{51.72 $\pm$ 0.78}\\
& 20 & 112 & 227.66 $\pm$ 7.75 & 168.01 $\pm$ 5.33 & \underline{122.15 $\pm$ 3.34} & 128.81 $\pm$ 3.87 & 135.05 $\pm$ 3.66 & \textbf{104.70 $\pm$ 1.02} \\
\bottomrule
\end{tabular}
}

\label{tab:random-mappings}
\end{table}

\section{Circuit Definitions}

In this appendix section, we briefly describe the types of circuits included in the datasets used for the empirical study.

\noindent\textbf{Amplitude Estimation (ae).}
Amplitude estimation circuits estimate an unknown amplitude encoded in a quantum state using amplitude amplification and phase-estimation-like subroutines~\cite{brassard2002quantum}. In routing benchmarks, these circuits are useful because their controlled amplification and estimation structure induces repeated long-range two-qubit interactions, making them a challenging case for evaluating whether a router can preserve favorable logical-to-physical qubit placements across multiple time slices.

\noindent\textbf{Deutsch--Jozsa (dj).}
Deutsch--Jozsa circuits solve the oracle problem of determining whether a Boolean function is constant or balanced~\cite{deutsch1992rapid}. It is a well known textbook algorithm that falls under umbrella of the hidden subgroup problem for commutative group.

\noindent\textbf{Greenberger--Horne--Zeilinger (ghz).}
GHZ circuits~\cite{greenberger1989going} prepare highly entangled multipartite states, commonly written as $$\frac{1}{\sqrt{2}}(\ket{0}^{\otimes n}+\ket{1}^{\otimes n}).$$ Their preparation requires propagating entanglement across many qubits through a sequence of two-qubit gates, making them useful benchmarks for evaluating how routing methods handle structured, non-local entangling patterns under limited hardware connectivity.

\noindent\textbf{Graph State (graphstate).}
Graph-state circuits prepare entangled states associated with an underlying graph, where vertices correspond to qubits and edges indicate entangling operations~\cite{hein2006entanglement,raussendorf2001oneway}. This is a much broader extension of the previous case.

\noindent\textbf{Quantum Approximate Optimization Algorithm (qaoa).}
QAOA circuits are variational circuits for approximate combinatorial optimization, alternating between problem-dependent cost operators and mixing operators~\cite{farhi2014quantum}. This is arguably most frequently used algorithm for combinatorial optimization problems. Their cost Hamiltonian layers induce two-qubit interactions determined by the problem graph, making them useful benchmarks for evaluating whether a routing can align logical interaction structure with the hardware topology.

\noindent\textbf{Quantum Fourier Transform (qft).}
QFT circuits \cite{weinstein2001implementation} implement the quantum analogue of the discrete Fourier transform and are central subroutines in algorithms such as phase estimation, Shor's factoring  and many other versions of the hidden subgroup problem.

\noindent\textbf{Quantum Neural Network (qnn).}
QNN circuits are parameterized quantum circuits used as trainable models in quantum machine learning~\cite{mitarai2018quantum,benedetti2019parameterized}.

\noindent\textbf{Quantum Phase Estimation (qpe).}
QPE circuits estimate the eigenphase of a unitary operator and serve as a fundamental primitive in many quantum algorithms~\cite{cleve1998quantum}. Their structure can induce repeated interactions between phase and system registers, making them useful for evaluating whether routing methods preserve favorable placements across successive time slices.

\noindent\textbf{Real Amplitudes (realamp).}
Real-amplitudes circuits are hardware-efficient variational ansatz circuits composed of alternating rotation and entangling layers that prepare states with real-valued amplitudes~\cite{qiskitrealamplitudes}.

\noindent\textbf{$\mathbb{SU}(2)$ (su2).}
The $\mathbb{SU}(2)$ circuits are parameterized ansatz circuits based on layers of single-qubit operations spanning $\mathbb{SU}(2)$ together with entangling gates~\cite{qiskitefficientsu2}.

\noindent\textbf{Variational Quantum Eigensolver (vqe).}
VQE circuits prepare parameterized trial states whose parameters are optimized classically to minimize the expectation value of a Hamiltonian~\cite{peruzzo2014variational}. The idea is very similar to that of QAOA but with only one Hamiltonian. These circuits have many applications in such domains as quantum chemistry, materials science, and condensed matter.

\noindent\textbf{Two-local (twolocal).}
Two-local circuits are parameterized ansatz circuits consisting of alternating single-qubit rotation layers and two-qubit entangling layers~\cite{qiskittwolocal}.

\noindent\textbf{Traveling Salesperson Problem (tsp).}
TSP circuits encode instances of the traveling salesperson problem, often through Ising/QUBO formulations or QAOA-style optimization circuits~\cite{lucas2014ising,farhi2014qaoa}.

\noindent\textbf{W State (wstate).}
W-state circuits prepare multipartite entangled states corresponding to an equal superposition over computational basis states with Hamming weight one~\cite{dur2000three}.

\section{Routing Performance on Different Types and Sizes of Quantum Circuits}

\subsection{Across Circuit Types}
\begin{figure}
    \centering
    \includegraphics[width=\linewidth]{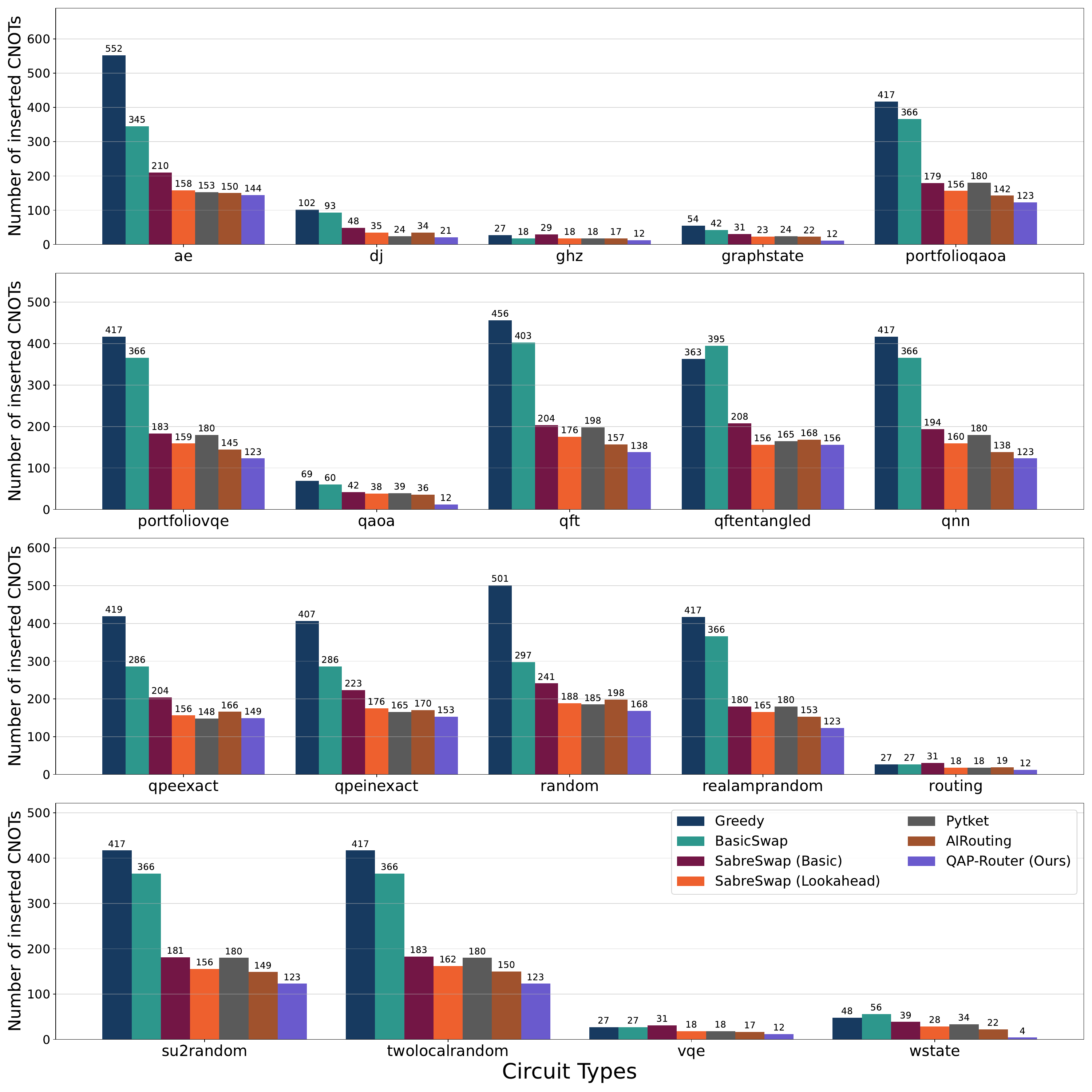}
    \caption{Comparison on the number of inserted CNOT gates between \texttt{QAP-Router} and state-of-the-art routers for 12-qubit circuits on 2D grid device. The results are shown across 19 circuit types in the MQTBench dataset.}
    \label{fig:12q-mqt-circuit-types}
\end{figure}

\begin{figure}
    \centering
    \includegraphics[width=\linewidth]{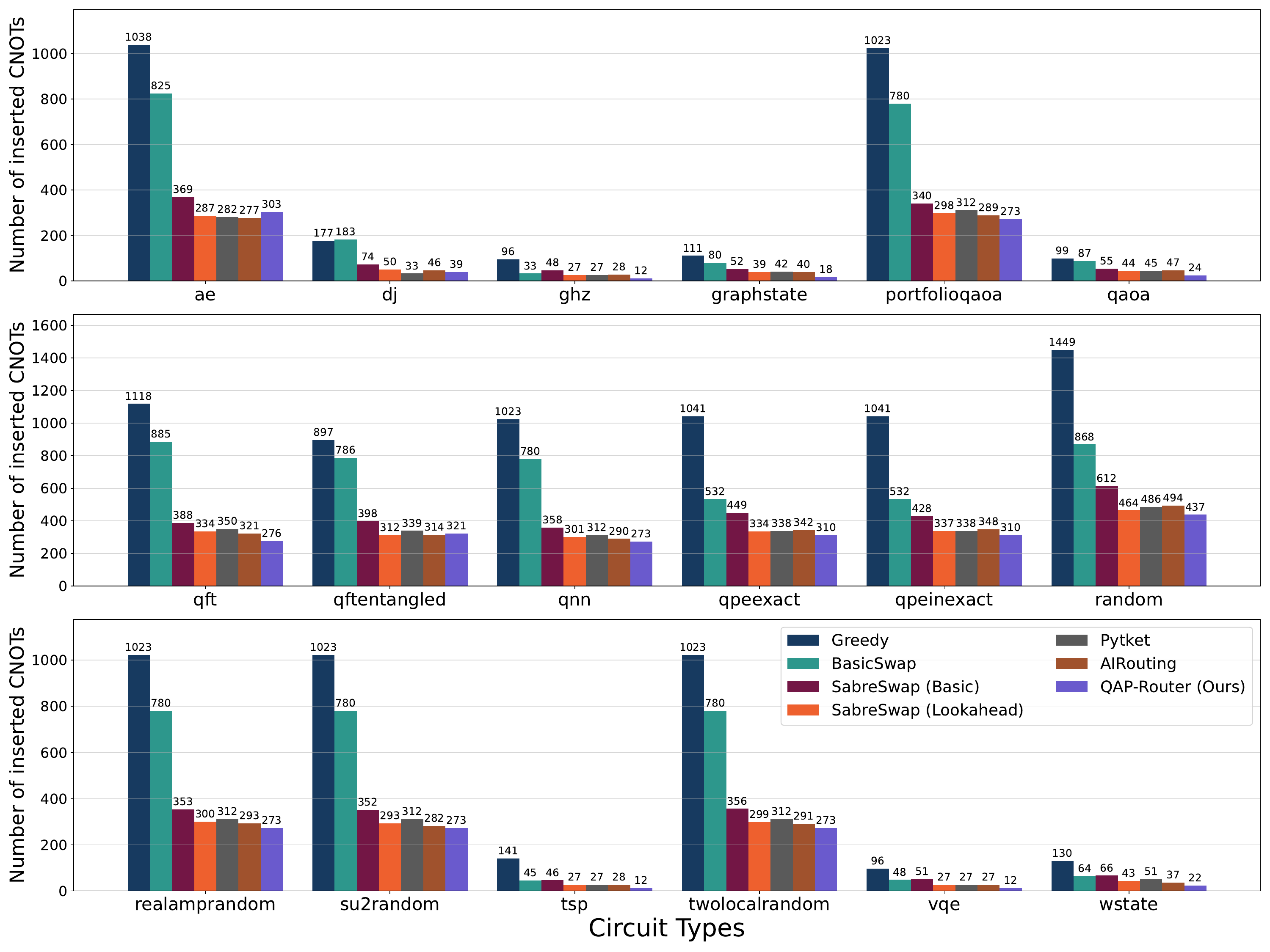}
    \caption{Comparison on the number of inserted CNOT gates between \texttt{QAP-Router} and state-of-the-art routers for 16-qubit circuits on 2D grid device. The results are shown across 18 circuit types in the MQTBench dataset.}
    \label{fig:16q-mqt-circuit-types}
\end{figure}

\begin{figure}
    \centering
    \includegraphics[width=\linewidth]{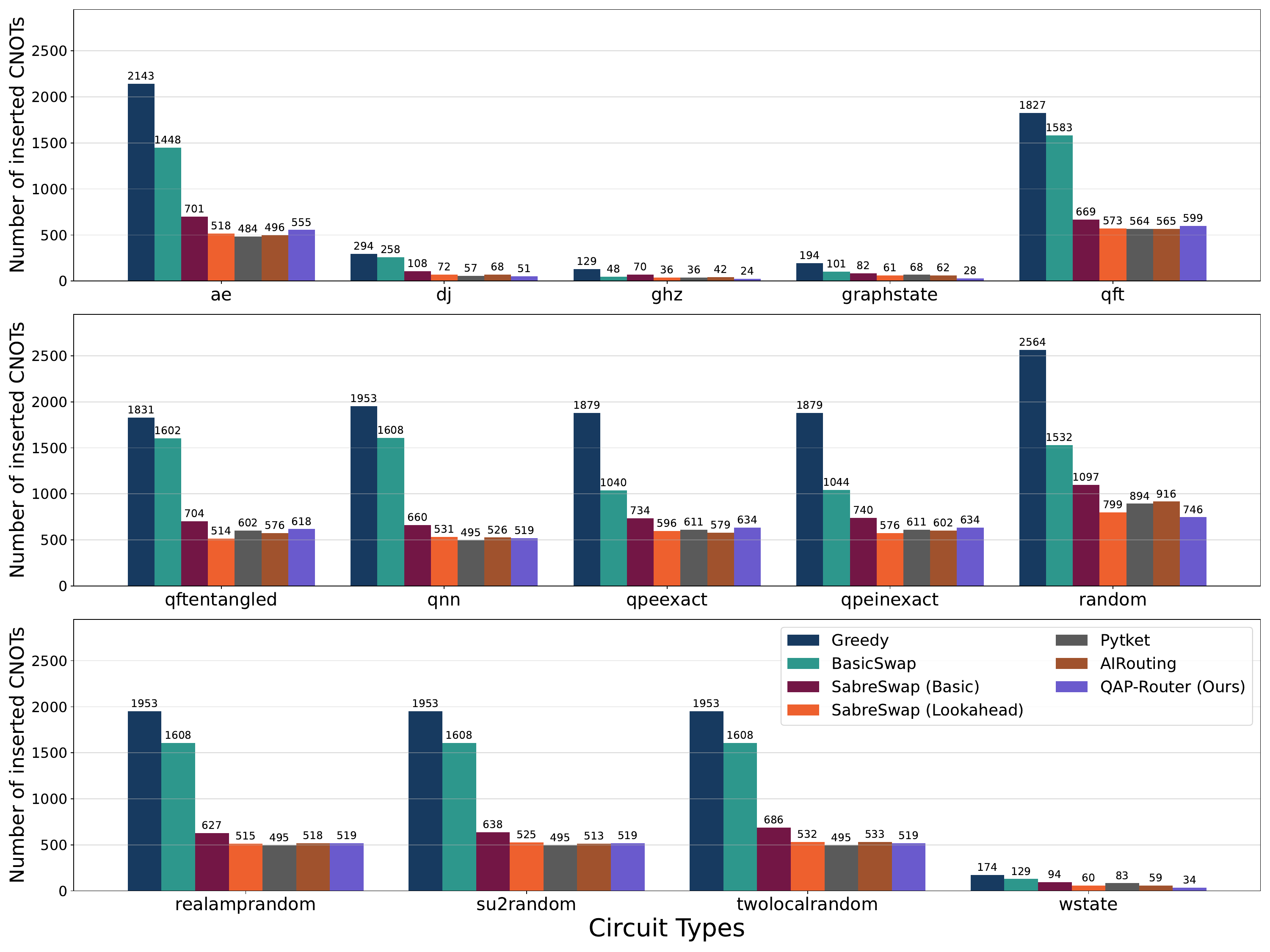}
    \caption{Comparison on the number of inserted CNOT gates between \texttt{QAP-Router} and state-of-the-art routers for 20-qubit circuits on 2D grid device. The results are shown across 14 circuit types in the MQTBench dataset.}
    \label{fig:20q-mqt-circuit-types}
\end{figure}

In Figures~\ref{fig:12q-mqt-circuit-types},~\ref{fig:16q-mqt-circuit-types} and~\ref{fig:20q-mqt-circuit-types}, we report the performance of each method across up to 19 circuit types from different quantum applications, such as quantum neural network, quantum approximate optimization algorithm, quantum variational eigensolver, etc.
We conduct the experiment on 2D Grid devices on the MQTBench dataset for 12-, 16-, and 20-qubit circuits.

Here, we take a closer look at the 20-qubit circuits where our method seems to struggle, as suggested by Table~\ref{tab:num-swaps}.
Circuits in the domains of Amplitude Estimation (ae), Quantum Fourier Transform (qft), and Quantum Phase Estimation (qpe) are where our method trails the strongest baselines most noticeably. 
These circuit families often contain highly structured long-range and repeated controlled interactions. 
Although the QAP objective provides a global topology-aware signal by jointly modeling logical interactions and physical distances, the current policy still makes sequential routing decisions over an evolving circuit state. 
Thus, for qft-, qpe-, and ae-like circuits, performance may depend not only on global spatial awareness, but also on accurately modeling long-horizon temporal structure across the circuit.

This indicates that the decay-weighted aggregation used to construct the effective flow matrix $\hat{\boldsymbol F}$ may be insufficient for larger-scale circuits, motivating more exploration in future work.
In contrast, methods such as \texttt{SabreSwap} may perform well on these regular circuit families due to their explicit bidirectional refinement.

\subsection{Across Circuit Sizes}
\begin{figure}
    \centering
    \includegraphics[width=0.85\linewidth]{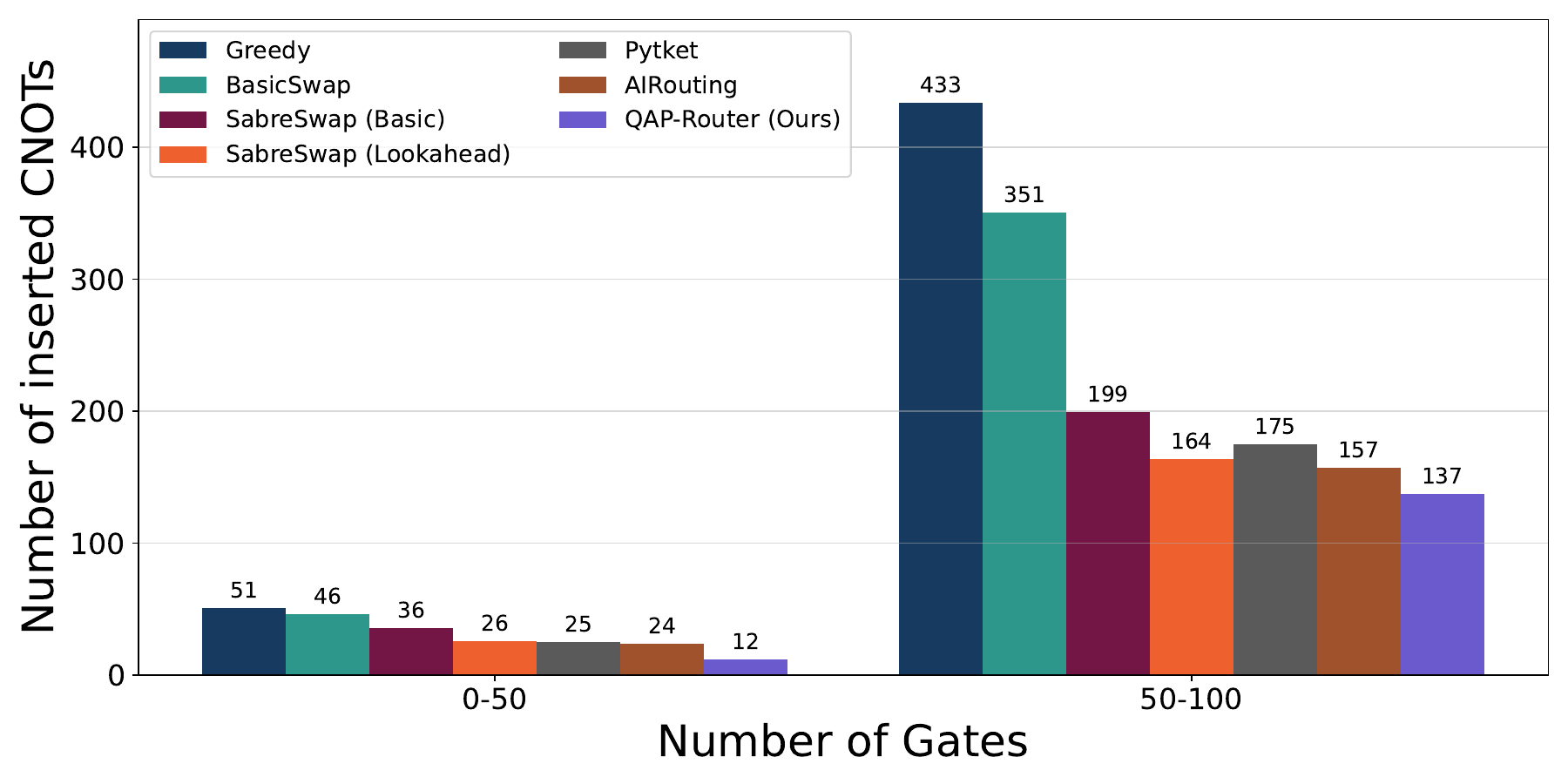}
    \caption{Comparison on the number of inserted CNOT gates between \texttt{QAP-Router} and state-of-the-art routers for 12 qubits. The results are shown across 2 ranges of gate numbers: 0-50 and 50-100, in the MQTBench dataset.}
    \label{fig:12q-gate}
\end{figure}

\begin{figure}
    \centering
    \includegraphics[width=0.85\linewidth]{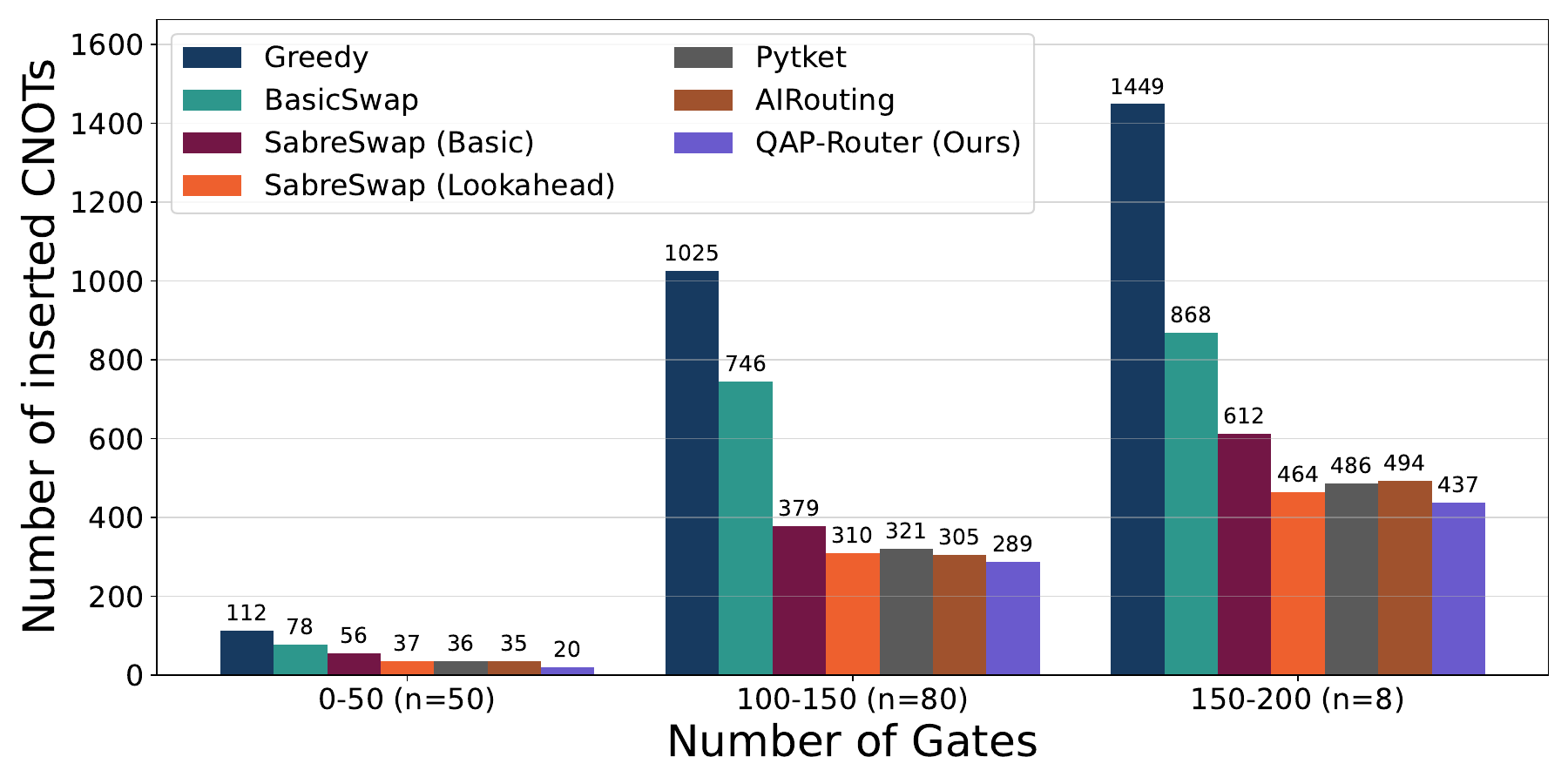}
    \caption{Comparison on the number of inserted CNOT gates between \texttt{QAP-Router} and state-of-the-art routers for 16 qubits. The results are shown across 3 gate ranges: 0-50, 100-150 and 150-200, in the MQTBench dataset.}
    \label{fig:16q-gate}
\end{figure}

\begin{figure}
    \centering
    \includegraphics[width=0.95\linewidth]{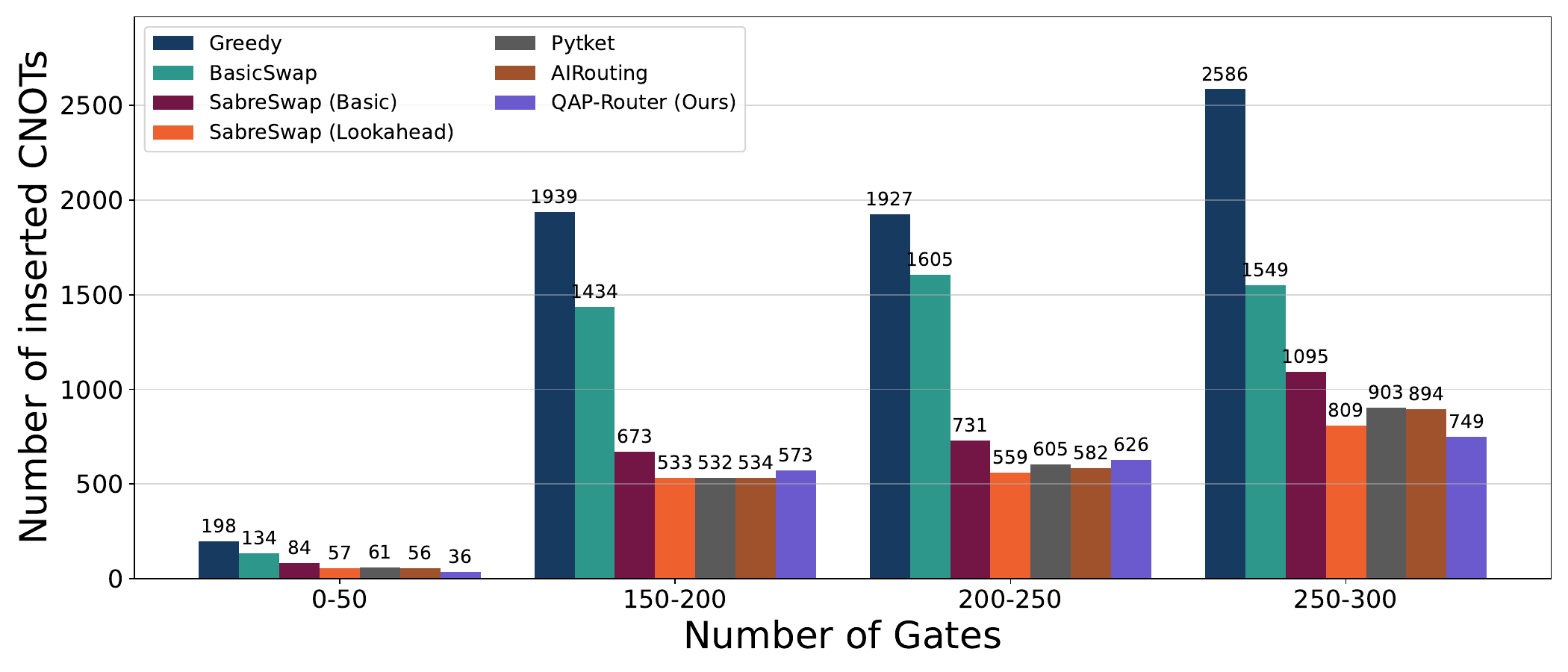}
    \caption{Comparison on the number of inserted CNOT gates between \texttt{QAP-Router} and state-of-the-art routers for 20 qubits. The results are shown across 4 gate ranges: 0-50, 150-200, 200-250, and 250-300, in the MQTBench dataset.}
    \label{fig:20q-gate}
\end{figure}

In Figures~\ref{fig:12q-gate},~\ref{fig:16q-gate} and~\ref{fig:20q-gate}, we report the performance of each method across several gate ranges. 
On 16 qubits, \texttt{QAP-Router} achieves the best results on all circuit sizes, reducing the average CNOT count by 16 relative to the second-best method, AIRouting, on medium-sized circuits. 
On larger circuits, \texttt{QAP-Router} ranks second, incurring an average overhead of 27 additional CNOT gates compared to \texttt{SabreSwap} with the lookahead heuristic.

\end{document}